\documentclass[12pt,preprint,prc,showpacs,preprintnumbers,amsmath,amssymb,floatfix]{revtex4}
\usepackage{amsfonts}
\usepackage{amssymb}
\usepackage{graphicx}
\usepackage{dcolumn}
\usepackage{subfigure}

\usepackage{amsmath}
\usepackage{amssymb}
\usepackage{dcolumn}
\usepackage{bm}
\usepackage{ulem} 
\usepackage[usenames]{color}
\usepackage{epsfig}
\usepackage{float}
\usepackage{multirow}
\usepackage{ifpdf}
\usepackage{bbm}
\usepackage{footnote}
\usepackage[version=3]{mhchem}

\allowdisplaybreaks[4]
\usepackage[colorlinks,
            linkcolor=blue,
            anchorcolor=blue,
            citecolor=blue,
            urlcolor=blue
            ]{hyperref}

\usepackage{CJK}

\begin{document}
\title{Nucleons pair shell model in M-scheme}
\begin{CJK*}{UTF8}{gbsn}
\author{B.C. He(何秉承)}
\affiliation{School of Physics, Nankai University, Tianjin, 300071, P.R. China}
\author{Y. Zhang(张宇)}
\email{dlzhangyu\_physics@163.com}
\affiliation{Department of Physics, Liaoning Normal University, Dalian 116029, P. R. China}
\author{Lei Li(李磊)}
\email{lilei@nankai.edu.cn}
\affiliation{School of Physics, Nankai University, Tianjin, 300071, P.R. China}
\author{Y.A. Luo(罗延安)}
\email{luoya@nankai.edu.cn}
\affiliation{School of Physics, Nankai University, Tianjin, 300071, P.R. China}
\author{F. Pan(潘峰)}
\affiliation{Department of Physics, Liaoning Normal University,
      Dalian 116029, P. R. China}
\affiliation{Department of Physics and Astronomy, Louisiana State
University, Baton Rouge, LA 70803, USA}
\author{J. P. Draayer}
\affiliation{Department of Physics and Astronomy, Louisiana State
University, Baton Rouge, LA 70803, USA}

\begin{abstract}
The nucleon pair shell model (NPSM) is casted into the so-called M-scheme for the cases with isospin symmetry and without isospin symmetry. The odd system and even system are treated on the same foot. The uncoupled commutators for nucleon-pairs, which are suitable for M-scheme, are given. Explicit formula of matrix elements in M-scheme for overlap, one-body operators, two-body operators are obtained. It is found that the $cpu$ time used in calculating the matrix elements in M-scheme is much shorter than that in the J-scheme of NPSM.
\end{abstract}

\maketitle

\end{CJK*}

\section{Introduction}

 Collective motions in nuclei, such as collective vibration, collective rotation, backbending phenomenon, giant resonances, etc. of medium and heavy nuclei, are extremely important.
 How to describe the collective motions of nucleus is a fundamental problem in the nuclear structure theory. Since nuclear shell model\cite{PhysRev.74.235,PhysRev.75.1766.2} includes all the  degrees of freedom, it can be used  to describe the collective phenomena technically\cite{RevModPhys.77.427}. With the development of the computer,
 the shell model Hamiltonian can be diagonalized in the model space  up to about $10^{10}$\cite{2013arXiv1310.5431S}. But even for medium and heavy nuclei, the shell model space is about $10^{14}-10^{18}$\cite{RevModPhys.59.339}, and the modern computer fail for all of these cases. Therefore, to apply the shell model theory to  medium and heavy nuclei, an efficient truncation scheme is necessary.
The interacting boson model (IBM)\cite{A2001The}  made a  great success  in nuclear structure theory\cite{PhysRevLett.92.212501,PhysRevLett.100.142501,PhysRevLett.93.212501,PhysRevLett.98.242502,HOU2010298,PhysRevC.95.061304}, in which the valence nucleons pairs are treated as $s$ bosons (with angular momentum J = 0) and $d$ bosons (with angular momentum J = 2). The vibrational spectrum, rotational spectrum and $\gamma$-unstable spectrum are corresponding to the $U(5)$, $SU(3)$ and $SO(6)$ limits in the IBM.

In 1993, a new technique, the generalized wick theory has been proposed to calculate the commutators for coupled operators and fermion clusters by Chen \textit{et al.}\cite{CHEN1993218,JINQUAN199361}. Based on this new technique, a nucleon pair shell model (NPSM) has been proposed\cite{CHEN1997686}, in which the building blocks of the configuration space are constructed by nucleon-pairs instead of the single valence nucleons.
Because of the success of the IBM, the shell model space was truncated to the
$SD$ pair subspace, which is the so-called SD-pair shell model(SDPSM). Previous works show that the collectivity of the low-lying states  can be described very well in the SDPSM\cite{LUO2000101,PhysRevC.62.014316,ZHAO20141, Zhao2000Validity}. The quantum phase transition and the properties of the critical point symmetry can also be reproduced very well in the SDPSM\cite{PhysRevC.80.014311}.
In 2000, new version of the NPSM was given by Zhao \textit{et al.}\cite{PhysRevC.62.014304}, in which the odd and even systems can be treated on the same foot. The NPSM was extended to include isospin symmetry in Ref. \cite{PhysRevC.87.044310,PhysRevC.87.044312}. The formalism in the NPSM with particle-hole coupling was also developed\cite{PhysRevC.97.024303}.
However, due to the $cpu$ time in calculating the matrix elements increase drastically with  the number of nucleon pairs, the maximum number of nucleon-pairs that the NPSM can handle is 5\cite{PhysRevC.77.047304} for identical nucleons  system. Therefore, an efficient method to calculate the matrix element is necessary in the NPSM.

In general, the shell model basis are constructed in the so-called J-scheme\cite{French1969} or the  M-scheme\cite{Whitehead1977}.
 Most of the large-scale shell model basis are constructed in M-scheme, since it does not need to calculate the 9j symbols and coefficients of fractional parentage\cite{2005RvMP...77..427C}. The old versions of the NPSM is constructed in J-scheme, and one has to re-couple and sum over all of the intermediate quantum numbers in calculating  the matrix elements. This procedure is too time consuming.  Because of the advantage of the M-scheme, it is interesting to cast the NPSM in the M-scheme, and this is the aim of this paper.

 The paper is organized as follows. In Sec.\ref{without_isospin}, the NPSM in M-scheme for the case without isospin symmetry are given; the NPSM in M-scheme for the case with isospin symmetry are presented in Sec.\ref{with_isospin}, and a brief summary and discussion  are given in Sec.\ref{summary}.

\section{The Hamiltonian, E2 transition operator and M1 transition operator}
As in the J-scheme, we still use a Hamiltonian consisting of the single-particle energy term
$H_0$,  and a residual interaction containing the multipole pairing between like nucleons
and the multipole-multipole interaction between all nucleons,

\begin{mathletters}

\begin{eqnarray}\label{eq4.1}
   & &  H = \sum_{\sigma=\pi, \nu} \left( H_{0}(\sigma) + V(\sigma) \right)
    + \sum_{t}\kappa_t Q^{t}_{\pi}\cdot Q^{t}_{\nu},
    \label{4.1a}\\
    & &  H_{0}(\sigma) = \sum_{a}  \epsilon_{a}\hat{n}_{a}, \quad
    V(\sigma) = \sum_{s} G_{s\sigma} A^{s\dag} \cdot A^s + \sum_t k_{t\sigma} Q^t \cdot Q^t, \nonumber \\
&& Q^{t} =  \sum_{i=1}^{n}(r_{i})^{t}Y_{t}(\theta_{i}\phi_{i}),  \nonumber
\end{eqnarray}
\end{mathletters}
where  $\epsilon_{a}$   and  $\hat{n}_{a}$ are the single-particle energy
 and the number operator respectively, and the pair creation operator is
 \begin{equation}
  A^{s\dagger}_{\nu} = \sum_{cd} y_{0}(cds) \left(
C^{\dagger}_{c} \times  C^{\dagger}_{d} \right)^{s}_{\nu}.\label{4.6a}
\end{equation}
Notice that the structure coefficients $y_{0}(cds)$  depend on the
Hamiltonian to be used and are in general different
from those in the building blocks, $y(cds)$, in Eq.\ (\ref{A_daggeroperator}).

The second quantized form of $Q^{t}$ is given by
 Eq.(\ref{Q_operator}) with the coefficients $q(cdt)$ equal to
\begin{eqnarray}
       q(cdt) &= & (-)^{c-\frac{1}{2}}
         \frac{\widehat{c}\widehat{d}}{\sqrt{20\pi}}
         C_{c \frac{1}{2}, d -\frac{1}{2}}^{t~0}   \Delta _{cdt}
         \langle Nl_{c} |r^{t}|Nl_{d} \rangle ,
\label{4.4b}   \\
\Delta_{cdt} & = & \mbox{$\frac{1}{2}$}\left[1+(-)^{l_{c}+l_{d}+t}\right],             \nonumber
\end{eqnarray}
where $N$ is the principal quantum number of the harmonic oscillator
 wave function, such that the energy is $(N+3/2)\hbar\omega_{0}$
and $l_{c}$  and $l_{d}$ are the orbital angular momentum of the s. p. levels
 $c$  and $d$, respectively.

The general form of the two-body realistic interaction in shell model as shown in the following can also use in this 
algorithm.
\begin{eqnarray}
V= \sum_{JT}\sum_{j_1 \leq j_2,~ j_3 \leq j_4}\frac{V_{JT}(j_1 j_2 j_3 j_4)}{\sqrt{1+\delta_{j_1 j_2}}\sqrt{1+\delta_{j_3 j_4}}} \left( A^{JT\dagger}(j_1 j_2)  \times A^{JT}(j_3 j_4) \right)^0
\end{eqnarray}

The $E2$ and $M1$ transition operators are
\begin{eqnarray}
E2 &=& e_{\pi}Q^2_{\pi} + e_{\nu} Q^2_{\nu}, \nonumber  \\
       T(M1) & =& T(M1)_\pi + T(M1)_\nu,~~~
       T(M1) = \sqrt{\frac{3}{4\pi}} (g_{l} {\bf L}+ g_{s} {\bf S}),
\end{eqnarray}
where $e_\pi$ and $e_\nu$ are effective charge of the protons and neutrons,
while   $g_{l}$ and $g_{s}$ are the orbital     and spin
effective gyro-magnetic ratios.
The total orbital angular momentum operator ${\bf L}$ and
total spin ${\bf S}$ can be identified with
collective dipole operators,
\begin{eqnarray}
               L_{\sigma}  \equiv  Q^{1}_{\sigma}
  = \sum_{cd}q(cd1) P^{1}_{\sigma}(cd),  \nonumber \\
               S_{\sigma}  \equiv  Q^{'1}_{\sigma}
  = \sum_{cd}q^{'}(cd1) P^{1}_{\sigma}(cd)
\label{7.1b}
\end{eqnarray}
with
\begin{eqnarray}
   q(cd1)=  (-1)^{l+1/2+d} \sqrt{ \mbox{$       \frac{l(l+1)} {3} $} }
   \hat{c}\hat{d}\hat{l}
  \left\{
\begin{array}{ccc}
  c   & d & 1 \\
  l   & l & \mbox{$ \frac{1}{2} $}
\end{array}
\right\}, \nonumber \\
   q^{'}(cd1)=  (-1)^{l+1/2+c} \mbox{$ \frac{1}{\sqrt{2}} $}
   \hat{c}\hat{d}
  \left\{
\begin{array}{ccc}
  c   & d & 1 \\
  \mbox{$ \frac{1}{2} $}   & \mbox{$ \frac{1}{2} $} & l
\end{array}
\right\}. \label{7.1c}
\end{eqnarray}

\section{NPSM in M scheme without isospin symmetry}\label{without_isospin}
In this section, the uncoupled commutators,  matrix elements for one-body operator and two body operator for the case without isospin is given in M-scheme. The odd system and even system are treated on the same foot.
\subsection{Uncoupled commutators for nucleon pairs}
As in the old version of the NPSM, the collective nucleons pair with angular momentum $r$ and projection $m$, designated as $A_\nu^{r\dagger}$, is built
from many non-collective pairs $A_\nu^r(cd)^\dagger$ in the single-particle orbits $a$ and $b$ in one major shell,
\begin{eqnarray} \label{A_daggeroperator}
A^{r\dagger}_m  = \sum_{ab}y(abr)A^{r\dagger}_m(ab)
\end{eqnarray}
where y(abr) are structure coefficients satisfying the symmetry
\begin{eqnarray}
y(abr) = -(-)^{a+b+r}y(bar)
\end{eqnarray}
Non-collective pair $A^{r\dagger}_m (ab)$ is
\begin{eqnarray}
 A^{r\dagger}_m(ab)&=(C^{a\dagger} \times C^{b\dagger})^r_m
 &=\sum\limits_{m_a,m_b} C^{rm}_{a,m_a,b m_b} C^{a\dagger}_{m_a}  C^{b\dagger}_{m_b}
\end{eqnarray}
where $C^{j\dagger}_m$ is a single particle creation operator, which create a nucleon in $j$ orbit with projection $m$, and $C^{rm}_{a,m_a,b m_b}$ is a Clebsch-Gordan coefficient. The time-reversed form of the single annihilate operator $C^j_m$ is
\begin{eqnarray}
\tilde{C}^j_m = (-)^{j-m}{C}^j_m
\end{eqnarray}
The time-reversed form of a collective pair is
\begin{align}
\tilde{A}^r_m &= \sum\limits_{ab} y(abr) \tilde{A}(ab)  \\
&=-\sum\limits_{ab} y(abr) ( \tilde{C}^{a}\times \tilde{C}^{b})^r_m \nonumber
\end{align}

A multipole operator or one-body operator $Q^t_\sigma$ is denoted by
\begin{flalign}\label{Q_operator}
&Q^t_\sigma = \sum\limits_{cd}q(cdt)(C^{c\dagger}\times \tilde C^{d})^t_\sigma  \\
&q(cdt) = -(-)^{c+d} \times q(dct)    \nonumber
\end{flalign}

The coupled commutators between two collective pairs is denoted as
\begin{flalign}
[\tilde A^r,A^{s\dagger}]^{t}_\sigma = \sum\limits_{\alpha \beta}C^{t \sigma}_{r\alpha,s\beta}[\tilde A^r_\alpha, A^{s\dagger}_\beta]
\end{flalign}

 Some crucial coupled commutators in the NPSM, taken from Ref.\cite{CHEN1993218},  are listed in the following. The coupled commutator between colective pair annihilation operator and colective pair creation operator is given by
\begin{align}
\label{eq1}
&[\tilde A^r,A^{s\dagger}]^t_\sigma= 2\hat{r}\delta_{rs}\delta_{t0}\sum\limits_{ab}y(abr)y(abs) - P^t_\sigma
\end{align}
where $P^t_\sigma$ is a new one-body operator.
\begin{align}\label{new_one-body-operator}
P_\sigma^t = 4\hat{r}\hat{s}\sum_{abd}y(abr)y(bds)  \left\{
   \begin{aligned}
&r &s~~~&t \\
&d &a ~~~&b \end{aligned}
    \right  \}(C^{d\dagger}\times \tilde C^a)^t_\sigma
\end{align}
The coupled commutator for collective pair and multipole operator would be given by,
\begin{align}    \label{eq2}
&[\tilde A^r, Q^t]^{r^\prime}_{m^\prime} =\tilde{A}^{r^\prime}_{m^\prime} 
\end{align}
where $\tilde{A}^{r^\prime}_{m^\prime}$ is a new colective pair annihilation operator,  which is
\begin{align}
&\tilde{A}^{r^\prime}_{m^\prime} =\sum\limits_{ad} y^\prime (dar^\prime) \tilde{A}^{r^\prime}_{m^\prime} (da)\\
&y^\prime(dar^\prime) = z(dar^\prime)-(-)^{a+d+r^\prime}z(adr^\prime),\nonumber\\
&z(dar^\prime)= \hat{r}\hat{t}\sum\limits_b y(abr)q(bdt)\left\{
   \begin{aligned}
&r &t~~~&r^\prime \\
&d &a ~~~&b \end{aligned}
    \right  \}\nonumber
\end{align}
By using Eq.(\ref{eq1}) and (\ref{eq2}) the coupled double commutator can be obtained, 
\begin{align}    
\label{eq3}
&\bigg[ A^{r_i},\big[A^{r_k},A^{s\dagger} \big]^t\bigg]^{r^\prime_i}
= \mathbb{B}^{r_i^\prime} 
\end{align}
where $\mathbb{B}^{r_i^\prime} $ is a new collective pair, 
\begin{align}
&\mathbb{B}^{r_i^\prime} = \sum\limits_{aa^\prime}y^\prime(aa^\prime r^\prime_i) \tilde{A}^{r^\prime_i}(aa^\prime) \\
&y^\prime(aa^\prime r^\prime_i) = z(aa^\prime r^\prime_i)-(-)^{a+a^\prime+r^\prime}z(a^\prime a r^\prime_i),\nonumber\\
&z(aa^\prime r^\prime_i)=-4\hat{r_i}\hat{r_k}\hat{s}\hat{t} \sum\limits_{b b^\prime}y(a^\prime b^\prime r_i)y(abr_k)y(bb^\prime s)\times \left\{
   \begin{aligned}
&r_k &s~~~&t \\
&a &b^\prime ~~~&b \end{aligned}
    \right  \}  \left\{
   \begin{aligned}
&r_i &t~~~&r^\prime \\
&a &a^\prime ~~~&b^\prime \end{aligned}
    \right  \} \nonumber
\end{align}
where $\hat t =\sqrt{2t+1}$. And the coupled commutator between single particle and one-body operator was given by eq.(2.11a) in Ref.\cite{PhysRevC.62.014304}, which is
\begin{align}
\label{eq4}
[\tilde C^j, Q^t]^{j^\prime}_{m^\prime}= (-)^{t-j-j^\prime}q(j,j^\prime,t)\frac{\hat t}{\hat j^\prime} \tilde C^{r^\prime}_{m^\prime}
\end{align}

The uncoupled commutator for nucleons pairs, which can be used to constructed the NPSM in M-scheme, are obtained  from the coupled commutators, which is
\begin{align}
&[A^r_\mu,A^{s\dagger}_\nu]   \\
=&(-)^{r-\mu} \sum\limits_{\alpha\beta}\delta_{-\mu\alpha}\delta_{\nu\beta}[\tilde A^r_\alpha,A^{s}_\beta]  \nonumber \\
=&(-)^{r-\mu}\sum\limits_{t\sigma}C^{t\sigma}_{r-\mu,s\nu} [\tilde{A}^r,A^{s\dagger}]^t_\sigma\nonumber
\end{align}

Based on Eqs.(\ref{eq1})-(\ref{eq4}), the uncoupled commutators in M-scheme can be obtained. The uncoupled commutator between colective pair annihilation operator and colective pair creation operator is given by
\begin{align}
&[A^r_\mu,A^{s\dagger}_\nu] \label{eq_pairs} \\
=&(-)^{r-\mu}\sum\limits_{t\sigma}C^{t\sigma}_{r-\mu,s\nu} [\tilde{A}^r,A^{s\dagger}]^t_\sigma\nonumber\\
=&2\delta_{r,s}\delta_{\mu,\nu}\sum_{ab}y(abr)y(abr)-(-)^{r-\mu}\sum\limits_{t\sigma}C^{t\sigma}_{r\mu,s\nu}P^t_\sigma\nonumber
\end{align}
where $P^t_\sigma$ is a new one-body operator, which is given in Eq.(\ref{new_one-body-operator}). The uncouple commutator for collective pair and one-body operator would be given by,
\begin{align}
&[A^r_m , Q^{t}_\sigma]  \label{eq_onebody} \\
=&(-)^{r+m}\sum\limits_{r^\prime m^\prime}C^{r^\prime m^\prime}_{r~-m,t\sigma} [\tilde{A}^r,Q^t]^{r^\prime}_{m^\prime}\nonumber\\
=&\sum\limits_{r^\prime m^\prime}\mathbb A^{r^\prime}_{-m^\prime} \nonumber
\end{align}
where $\mathbb A^{r^\prime}_{-m^\prime}$ is a new collective pair, which is
\begin{align}
&\mathbb A^{r^\prime}_{-m^\prime} = \sum_{ad} y^\prime(dar^\prime) A^{r^\prime}_{-m^\prime}(da) \\
&y^\prime(dar^\prime) = z(dar^\prime)-(-)^{a+d+r^\prime}(adr^\prime), \nonumber\\
z(dar^\prime)=& \hat{r}\hat{t}\sum\limits_{r^\prime m^\prime} (-)^{r+r^\prime+m- m^\prime} C^{r^\prime m^\prime}_{r~-m,t\sigma}\sum\limits_{b}y(abr)y(bdt) \left\{
   \begin{aligned}
&r &t~~~&r^\prime \\
&d &a ~~~&b \end{aligned}
    \right  \}  \nonumber
\end{align}
By using Eq.(\ref{eq_pairs}) and (\ref{eq_onebody}) the uncoupled double commutator can be obtained, 
\begin{align}
&\bigg[ A^{r_i}_{m_i} ,\big[A^{r_k}_{m_k} ,A^{s\dagger}_m \big]\bigg] \label{eq_double} \\
=& (-)^{r_{k} + r_{i} - m_k - m_i } \sum\limits_{
\substack{
t\sigma \\
r^\prime m^\prime}}
C^{r^\prime m^\prime}_{r_i -m_i ,t \sigma} C^{t\sigma}_{r_k -m_k,s m}  \bigg[ \tilde{A}^{r_i},\big[\tilde{A}^{r_k},A^{s\dagger}\big]^{t}\bigg]^{r^\prime}_{m^\prime}  \nonumber\\
=&\sum\limits_{r^\prime m^\prime} \mathbb B^{r^\prime}_{-m^\prime}\nonumber 
\end{align}
where $\mathbb B^{r^\prime}_{-m^\prime}$ is a new collective pair, which is 
\begin{align}
&\mathbb B^{r^\prime}_{-m^\prime} = \sum\limits_{
aa^\prime}y^\prime(aa^\prime r^\prime_i) A^{r^\prime}_{-m^\prime}(aa^\prime)\\
&y^\prime(aa^\prime r^\prime_i) = z(aa^\prime r^\prime_i)-(-)^{a+a^\prime+r^\prime}z(a^\prime a r^\prime_i),\nonumber\\
&z(aa^\prime r^\prime_i)=-4\hat{r_i}\hat{r_k}\hat{s}\sum\limits_{
t\sigma} \hat{t} (-)^{r_{k} + r_{i} + r^\prime - m } C^{r^\prime m^\prime}_{r_i -m_i ,t \sigma} C^{t\sigma}_{r_k -m_k,s m} \sum\limits_{
b b^\prime}y(a^\prime b^\prime r_i)y(abr_k)y(bb^\prime s)\nonumber\\ &~~~~~\times \left\{
   \begin{aligned}
&r_k &s~~~&t \\
&a &b^\prime ~~~&b \end{aligned}
    \right  \}  \left\{
   \begin{aligned}
&r_i &t~~~&r^\prime \\
&a &a^\prime ~~~&b^\prime \end{aligned}
    \right  \}  \nonumber
\end{align}
By using Eq.(\ref{eq_onebody}) recursively, the uncoupled double commutator between pair annihilation operator and multipole-multipole interaction operator can be obtained, which is
\begin{align}
&\sum\limits_{\sigma}(-)^\sigma  \bigg [ \big[A^{r}_{m},  Q^t_\sigma \big], Q^t_{-\sigma} \bigg]     \label{eq_pair_m-m}  \\
=&\sum\limits_{r^\prime}(-)^{r-r^\prime}\frac{\hat r^\prime}{ \hat r } \mathbb{A}^r_m \nonumber
\end{align}
where $\mathbb{A}^r_m$ is a new collective pair, which is given by,
\begin{align}
&\mathbb{A}^r_m =\sum\limits_{ab} y^\prime (abr)  A^r_m(ab) \\
&y^\prime (abr) =  \big[ h(abr)-(-)^{a+b+r}h(bar)\big]  \nonumber\\
&h(abr) = 2 \hat r^\prime \hat r (2t+1) \sum\limits_{dd_0}y(d d_0 r) q(d_0 b t) q(dat) \left\{
   \begin{aligned}
&r &t~~~&r^\prime \\
&b &d~~~&d_0 \end{aligned}
    \right  \}  \left\{
   \begin{aligned}
&r &t~~~&r^\prime \\
&d &b~~~&a \end{aligned}
    \right  \} \nonumber
\end{align}

The uncoupled commutator between single nucleon annihilation operator and one-body operator can be expressed as,
\begin{align}
 &[C^j_{m_0}, Q^t_\sigma]     \label{eq_single}  \\
 =&-\sum\limits_{j^\prime m^\prime}(-)^{t-\sigma} C^{j^\prime m^\prime}_{j~-m_0,t\sigma}\frac{\hat t}{ \hat{j^\prime} }q(jj^\prime t) C^{j^\prime}_{-m^\prime}\nonumber
\end{align}
And the uncoupled double commutator for single nucleon can also be obtained, which is
\begin{align}
 &\bigg [C^j_{m_0}, \big[ A^{r_k}_{m_k},A^{s\dagger}_{m}  \big]\bigg]     \label{eq_single_double}  \\
 =&\sum\limits_{\substack{
t\sigma \\
r^\prime m^\prime}}(-)^{r_k+j-m_k+m_0} C^{t\sigma}_{r_k~-m_k,sm} C^{r^\prime m^\prime}_{j~-m_0,t\sigma} \bigg [C^j, \big[ A^{r_k},A^{s\dagger} \big]^t\bigg]^{r^\prime}_{m^\prime}  \nonumber\\
 =&4\sum\limits_{\substack{
t\sigma \\
r^\prime m^\prime}}(-)^{r_k+t-m}  C^{t\sigma}_{r_k~-m_k,sm} C^{r^\prime m^\prime}_{j~-m_0,t\sigma}  \frac{\hat r_k \hat s \hat t}{\hat r^\prime}  \sum\limits_{b} y(r^\prime b r_k) y(b j s) \left\{
   \begin{aligned}
&r_k &s~~~&t \\
&j &r^\prime ~~~&b \end{aligned}
    \right  \} C^{r^\prime}_{-m^\prime} \nonumber
\end{align}
By recursive applications of Eq.(\ref{eq_single}), we can obtain the uncoupled double commutator between single nucleon operator and multipole-multipole interaction operator, which is
\begin{align}
 &\sum\limits_{\sigma}(-)^\sigma  \bigg [ \big[C^j_{m_0},  Q^t_\sigma \big], Q^t_{-\sigma} \bigg]     \label{eq_single_m-m}  \\
 =& \sum\limits_{j^\prime}(-)^{j-j^\prime}\frac{2t+1}{2j+1}q(jj^\prime t)q(j^\prime jt)C^j_{m_0}\nonumber
\end{align}

\subsection{commutators  in M-scheme}  
The odd system with $2N+1$ nucleons and even system with $2N$ nucleons are treated on the same foot. The creation operator coupled successively to the total angular
momentum projection $M$ is designated by
\begin{align}\label{basis_convention}
A^\dagger(r_0 m_0,\dots,r_N m_N)_M = A^{r_0\dagger}_{m_0}\cdot A^{r_1\dagger}_{m_1}\dots A^{r_N \dagger}_{m_N},~~M = \sum\limits_{i=0}^{N}m_i
\end{align}
with the convention that
\begin{align}
A^{r_0\dagger}_{m_0} = \left\{  \begin{aligned}
&1     &~~~\text{for even system,  }&m_0 \equiv 0 \\
&C^{r_0\dagger}_{m_0}  &~~~\text{for odd system,  } &r_0 \equiv j  \end{aligned} \right.
\end{align}
The annihilation operator $A_M$ is defined as following
\begin{align}
A(r_0 m_0,\dots,r_N m_N)_M = A^{r_0}_{m_0}\cdot A^{r_1}_{m_1}\dots A^{r_N }_{m_N},~~M = \sum\limits_{i=0}^{N}m_i
\end{align}
the convention of $A^{r_0}_{m_0}$ is similar to that of Eq.(\ref{basis_convention}).
Then one can get the commutator between the annihilation operator $A_M$  and pair creation operator, which is
\begin{align}
&[A(r_0 m_0,\dots,r_N m_N)_{M},A^{s \dagger}_m] \label{eq_basisandpair}  \\
=&\sum\limits_{k=1}^{N} \bigg[ \varphi \delta_{r_k,s}\delta_{m_k,m} A(r_0 m_0,\dots,r_{k-1} m_{k-1},r_{k+1} m_{k+1},\dots, r_N m_N)_{M-m}  \nonumber \\
+&\sum\limits_{i=k-1}^{0~or~1}\sum\limits_{~r^\prime_i m^\prime_i}  A(r_0 m_0,\dots, r_{i}^\prime m_{i}^\prime,\dots ,r_{k-1} m_{k-1},r_{k+1} m_{k+1},\dots, r_N m_N)_{M-m} \nonumber\\
+& \sum\limits_{t \sigma} P^t_\sigma \times A(r_0 m_0,\dots ,r_{k-1} m_{k-1},r_{k+1} m_{k+1},\dots, r_N m_N)_{M-m_k}  \bigg]\nonumber
\end{align}
where $\varphi=2 \sum\limits_{ab}y(a b r_k)y(a b s)$, the summation runs over $i $ is from k-1 to 0 or 1, corresponding to the odd  system or even  system respectively,  $A^{r^\prime_i}_{m^\prime_i}$ represents a new collective pair ($i\ne0$) or a single nucleon ($i=0$),
\begin{align}
A^{r^\prime_i}_{m^\prime_i} = \bigg[ A^{r_i}_{m_i} ,\big[A^{r_k}_{m_k} ,A^{s\dagger}_m \big]\bigg]
\end{align}
the explicit form of the new pair has already been given in Eq.(\ref{eq_double}) and (\ref{eq_single_double}), and $P^t_\sigma$ has been given in Eq.(\ref{eq_pairs}). Using Eq.(\ref{eq_basisandpair}), we have the commutation relation between pairing interaction and the annihilation operator $A_M$,
\begin{align}\label{eq_paring}
&[A(r_0 m_0,\dots,r_N m_N)_{M},A^{s \dagger}\cdot A^s]\\
=&\sum\limits_{k=1}^{N} \bigg[ \varphi \delta_{r_k,s}\delta_{m_k,m} A(r_0 m_0,\dots,s m_{k},\dots, r_N m_N)_{M}  \nonumber \\
+&\sum\limits_{i=k-1}^{0~or~1}\sum\limits_{~r^{\prime}_i m^{\prime}_i m}  A(r_0 m_0,\dots, r_{i}^\prime m_{i}^\prime,\dots ,r_{k-1} m_{k-1},r_{k+1} m_{k+1},\dots, r_N m_N, s m)_{M}  \nonumber\\
+&\sum\limits_{t \sigma m} P^t_\sigma \times A(r_0 m_0,\dots ,r_{k-1} m_{k-1},r_{k+1} m_{k+1},\dots, r_N m_N, s m)_{M-m_k+m} \bigg] \nonumber
\end{align}

The commutator between one body operator and the annihilation operator $A_M$ is obtained,
\begin{align}\label{eq_basis_onebody}
&[A(r_0 m_0,\dots,r_N m_N)_{M},Q^{t}_\sigma]  \\
=&\sum\limits_{k=N}^{0~or~1} \sum\limits_{r^\prime_k m^\prime_k} A(r_0 m_0, r_1 m_1,\dots,r^\prime_k m^\prime_k,\dots,r_N m_N)_{M-\sigma} \nonumber
\end{align}
where summation runs over $k$ is from $N$ to $0$ or $1$, corresponding to odd  system or even  system, respectively, and $A^{r^\prime_k}_{m^\prime_k}$ represents a new collective pair ($k\ne0$) or a single nucleon ($k=0$),
\begin{align}
A^{r^\prime_k}_{m^\prime_k} = [A^{r_k}_{m_k}, Q^t_\sigma]
\end{align}
the explicit forms of the new pair have already been given in Eq.(\ref{eq_onebody}) and (\ref{eq_single}).

The commutator between multipole-multipole operator and the annihilation operator $A_M$ is
\begin{align}\label{eq_basis_m-m}
&[A(r_0 m_0,\dots,r_N m_N)_{M}, \sum\limits_{\sigma} (-)^\sigma Q^{t}_\sigma  Q^{t}_{-\sigma} ] \\
=&\sum\limits_{k=N}^{0~or~1}\bigg[ A(r_0 m_0, r_1 m_1,\dots,(r_k m_k)_B,\dots,r_N m_N)_{M}\nonumber\\
+&  \sum\limits_{i=k-1}^{0~or~1}  \sum\limits_{\substack{
r_i^\prime m_i^\prime\\
r_k^\prime m_k^\prime}}  \sum\limits_{\sigma} (-)^{\sigma}    A(r_0 m_0, r_1 m_1,\dots,r^\prime_i m^\prime_i,\dots,r^\prime_k m^\prime_k,\dots,r_N m_N)_{M}  \nonumber\\
+& \sum\limits_{\sigma} \sum\limits_{
r_k^\prime m_k^\prime}   (-)^{\sigma} Q^t_{-\sigma} \times A(r_0 m_0, r_1 m_1,\dots,r^\prime_k m^\prime_k,\dots,r_N m_N)_{M-\sigma} \bigg] \nonumber
\end{align}
where summation range over $k (i)$ is from $N (k-1)$ to $0$ or $1$, corresponding to odd  system or even  system respectively, and $r^\prime_k m^\prime_k (r^\prime_i m^\prime_i)$  represent for a new collective pair ($k\ne0$) or a single nucleon ($k=0$),
\begin{align}
A^{r^\prime_k}_{m^\prime_k} = [A^{r_k}_{m_k}, Q^t_\sigma] \\
A^{r^\prime_i}_{m^\prime_i} = [A^{r_i}_{m_i}, Q^t_{-\sigma}]\nonumber 
\end{align}
the explicit form of new pair have already given in Eq.(\ref{eq_onebody}) and (\ref{eq_single}). $(r_k m_k)_B$ denote a new collective pair ($k\ne0$) or a single nucleon ($k=0$) obtained by uncoupled double commutator,
\begin{align}
&\big(A^{r_k}_{m_k})_B = \sum\limits_{\sigma}(-)^\sigma  \bigg [ \big[A^{r}_{m},  Q^t_\sigma \big], Q^t_{-\sigma} \bigg]     \label{}  \\
&(A^{r_0}_{m_0})_B =\bigg [C^j_{m_0}, \big[ A^{r_k}_{m_k},A^{s\dagger}_{m}  \big]\bigg]  \nonumber
\end{align}
the explicit results have been given in Eq.(\ref{eq_pair_m-m}) and (\ref{eq_single_m-m}).

\subsection{Matrix elements of overlap and interactions}
An $N$-pair or $N$-pair plus one single nucleon state in m-scheme is designated as
\begin{align}
|\alpha,M_N> &= |r_0m_0,r_1m_1,\dots r_Nm_N;M_N\rangle \\
&\equiv A^\dagger(r_0m_0,r_1m_1,\dots r_Nm_N)|0\rangle \nonumber
\end{align}
where $r_0m_0$ represents $1$ in even particle number system or a single nucleon in odd particle number system, $\alpha$ denotes the additional quantum numbers
\begin{align}
\alpha =(r_0m_0,\dots,r_N,m_N)
\end{align}
It is interesting to note that $\alpha$ is redundant, since it's already been included in the total projection number $M$.

For proton-neutron coupled system, the basis are constructed by coupling the protons and neutrons to the state with total projection number $M$,
\begin{align}
 |\alpha ,M_n M_p;M\rangle =|\alpha_p,M_p\rangle|\alpha_n,M_n\rangle
\end{align}

The overlap between two states is a key quantity, since the matrix elements of one-body and two-body interaction can all be expressed as a summation of the overlaps. From Eq.(\ref{eq_basisandpair}), we can get the overlap between two states,
\begin{align}\label{eq_overlap}
&\langle 0|A_{M_1}A^{\dagger}_{M_2}|0\rangle\equiv  \langle r_0\mu_0,r_1\mu_1,\dots r_N\mu_N;M_1|s_0\nu_0,s_1\nu_1,\dots, s_N\nu_N;M_2\rangle \\\
=&\sum\limits_{k=1}^{N} \bigg[ 2\sum_{ab}y(abr_k)y(abs_N) \delta_{r_k,s_N}\delta_{\mu_k,\nu_N} \nonumber \\
&\times \langle r_0 \mu_0\dots r_{k-1} \mu_{k-1},r_{k+1} \mu_{k+1}\dots r_N \mu_N;M_1-\nu_N|s_0\nu_0\dots s_{N-1}\nu_{N-1};M_2-\nu_N \rangle  \nonumber \\
&+\sum\limits_{i=k-1}^{0~or~1}\sum\limits_{~r^\prime_i \mu^\prime_i}
\nonumber \\ &\times  \langle r_0 \mu_0\dots r^\prime_{i} \mu^\prime_{i}\dots r_{k-1} \mu_{k-1},r_{k+1} \mu_{k+1}\dots r_N \mu_N;M_1-\nu_N|s_0\nu_0\dots s_{N-1}\nu_{N-1};M_2-\nu_N\rangle \bigg] \nonumber
\end{align}
One can see that although the overlap is still calculated recursively, the most time consuming factor, the re-coupling of the angular momentum, is not needed.
The summation over the projection $\mu^\prime_i$ of the new pair is redundant, since it is  a constant value $\mu_i+\mu_k-\nu_N$. The  overlap  for one pair state in M-scheme is same as that  in J-scheme, which is
\begin{align}
&\langle r_1 \mu_1 | s_1 \nu_1 \rangle
=2\delta_{r_1,s_1} \delta_{\mu_1,\nu_1} \sum\limits_{ab} y(abr_1)y(abs_1)
\end{align}
The overlap for  one pair plus one single nucleon is given as following,
\begin{align}
&\langle r_0\mu_0,r_1 \mu_1 | s_0\nu_0 ,s_1 \nu_1 \rangle  \\
=&2\delta_{r_1,s_1} \delta_{\mu_1,\nu_1} \delta_{r_0,s_0} \delta_{\mu_0,\nu_0} \sum\limits_{ab} y(abr_1)y(abs_1) \nonumber \\
&+4\hat r_1 \hat s_1\sum\limits_{JM}C^{JM}_{r_0 \mu_0,r_1 \mu_1}C^{JM}_{s_0 \nu_0,s_1 \nu_1} \sum\limits_{a}y(a s_0 r_1) (a r_0 s_1) \left\{
   \begin{aligned}
&s_1 &r_0~~~&a \\
&r_1 &s_0 ~~~&J \end{aligned}
    \right  \}   \nonumber
\end{align}
where $r_0(s_0)$ denote the single nucleon, and the summation over projection M is redundant, and it should be $\mu_0 + \mu_1$.

The matrix elements of a pair creation operator $A^{s\dagger}_\nu$ between two states differing by one pair is equal to an overlap,
\begin{align}
&\langle r_0 \mu_0,\dots,r_N \mu_N|A^{s\dagger}_\nu | s_0 \nu_0,\dots,s_{N-1}\nu_{N-1} \rangle  \\
=& \langle r_0 \mu_0,\dots,r_N \mu_N| s_0 \nu_0,\dots,s_{N-1}\nu_{N-1},s\nu \rangle \nonumber
\end{align}

Using Eq.(\ref{eq_paring}),  the matrix elements for the pairing interaction can be written as
\begin{align} \label{eq_paring_interaction}
 &\langle r_0\mu_0,r_1\mu_1,\dots r_N\mu_N;M|A^{s\dagger}\cdot A^{s}|s_0\nu_0,s_1\nu_1,\dots, s_N\nu_N;M \rangle \\
=&\sum\limits_{k=1}^{N} \bigg[ \varphi \delta_{r_k,s}\delta_{\mu_k,\nu} \langle r_0 \mu_0\dots \boldsymbol{s} \boldsymbol{\mu}_{k}\dots r_N \mu_N;M|s_0\nu_0\dots s_{N}\nu_{N};M\rangle  \nonumber \\
+&\sum\limits_{i=k-1}^{0~or~1}\sum\limits_{~r^\prime_i \mu^\prime_i m}  \langle r_0 \mu_0\dots \boldsymbol {r}^\prime_i \boldsymbol{\mu}^\prime_i\dots \boldsymbol{s} \boldsymbol{m}  \dots r_N \mu_N;M|s_0\nu_0\dots s_{N}\nu_{N};M\rangle \bigg] \nonumber
\end{align}
where the summation over the projection $m$ in the second term represents  the projection of pairing interacting $\sum\limits_{m}A^{s\dagger}_m A^{s}_m$, and the summation over the projection $\mu^\prime_i$ of the new pair is redundant, it should be a constant value $\mu_i+\mu_k-m$.

By using Eq.(\ref{eq_basis_onebody}),  the  matrix element of one-body operator can be written as
\begin{align}
& \langle r_0 \mu_0,\dots,r_N \mu_N ;{M_1}| Q^{t}_\sigma | s_0\nu_0,\dots, s_N\nu_N;M_2  \rangle  \\
=&\sum\limits_{k=N}^{0~or~1}\sum\limits_{r^\prime_k ,\mu^\prime_k}\langle r_0 \mu_0,\dots,\boldsymbol{r}^\prime_k \boldsymbol{\mu}^\prime_k,\dots,r_N \mu_N;M_1-\sigma|s_0\nu_0,\dots, s_N\nu_N;M_2 \rangle \nonumber
\end{align}
where the summation over the projection $\mu^\prime_k$ of new pair is redundant, and it should be a constant value $\mu_k-\sigma$.

The multipole-multipole interaction is given by
\begin{align}
Q^t\cdot Q^t =\sum\limits_{\sigma} (-)^\sigma Q^{t}_\sigma ~ Q^{t}_{-\sigma}
\end{align}
By using Eq.(\ref{eq_basis_m-m}),  the matrix elements of the multipole-multipole interaction between like nucleons is
\begin{align} \label{eq_QQ_interaction}
& \langle r_0 \mu_0,\dots,r_N \mu_N ;{M}| Q^{t} \cdot Q^{t} | s_0\nu_0,\dots, s_N\nu_N;M  \rangle \\
=&\sum\limits_{k=N}^{0~or~1}\bigg[  \langle r_0 m_0,\dots,(r_k m_k)_B,\dots,r_N m_N;M |  s_0\nu_0,\dots, s_N\nu_N;M \rangle \nonumber   \\
+&  \sum\limits_{i=k-1}^{0~or~1}  \sum\limits_{\substack{
r_i^\prime m_i^\prime\\
r_k^\prime m_k^\prime}}  \sum\limits_{\sigma} (-)^{\sigma} \langle r_0 m_0,\dots,r^\prime_i m^\prime_i,\dots,r^\prime_k m^\prime_k,\dots,r_N m_N;M|s_0\nu_0,\dots, s_N\nu_N;M \rangle  \bigg] \nonumber
\end{align}
where the summation over the projection $\mu^\prime_k~(\mu^\prime_i)$ of new pair is redundant, and it should be a constant value $\mu_k-\sigma~(\mu_k+\sigma)$.

The matrix elements of multipole-multipole interaction between proton-neutron  can be expressed of the product of  matrix elements of the multipole operator for protons and neutrons,
\begin{align}
&\langle M_n,M_p;M | Q^t(\nu)\cdot Q^t(\pi)| M_n^\prime,M_p^\prime;M \rangle \\
=&\sum\limits_{\sigma}(-)^\sigma \langle \alpha_p,M_p|  Q^t_\sigma(\pi)| \alpha^\prime_p,M^\prime_p\rangle      \langle \alpha_n,M_n|  Q^t_{-\sigma}(\nu)| \alpha^\prime_n,M^\prime_n\rangle  \nonumber
\end{align}

\subsection{Diagonalization of the Hamiltonian}
In this part, a general Hamiltonian are considered, which consist of the single-particle energy term $H_0$, and a residual interaction containing $H(\sigma = \pi ~or~ \nu)$ for identical particles and interaction between proton and neutron $H_{\pi \nu}$, 
\begin{align}
&H = \sum\limits_{\sigma = \pi,\nu} ( H_0 (\sigma) + H(\sigma) ) + H_{\pi \nu}, \\
&H_0(\sigma) = \sum_a \epsilon_{a,\sigma} \hat n_{a,\sigma},\nonumber
\end{align}
where $\epsilon_{a,\sigma}$ and $\hat n_{a,\sigma}$ are single-particle energy and the number operator respectively. The Hamiltonian can be diagonialized either in the non-orthonormal basis $|\alpha_n, M_n \alpha_p M_p; M\rangle$, or the orthonormal basis $|\xi_n M_n,~ \xi_p M_p; M\rangle$. For both neutron and proton basis, the orthonormal basis $|\xi_\sigma M_\sigma\rangle$ ($\sigma = \pi$ or $\nu$) can be expanded in terms of the non-orthonormal basis  $|\alpha_\sigma M_\sigma\rangle$,
\begin{eqnarray}\label{eq_orthonormal}
|\xi_\sigma M_\sigma\rangle = \sum_{\alpha_\sigma} Z^{M_\sigma}_{\xi_\sigma, \alpha_\sigma}|\alpha_\sigma M_\sigma\rangle
\end{eqnarray}
where the coefficients $Z^{M_\sigma}_{\xi_\sigma, \alpha_\sigma}$ are found in the following way. Suppose that $\bar {\textbf{Z}}^{M_\sigma}_{\xi_\sigma}$ is an orthonormalized eigenvector of the overlap matrix $\Lambda^{M_\sigma}$,
\begin{eqnarray}
\Lambda^{M_\sigma}_{\alpha^\prime, \alpha} = \langle \alpha^\prime M_\sigma| \alpha M_\sigma\rangle
\end{eqnarray}
corresponding to the eigenvalue $\lambda_{M_\sigma, \xi_\sigma}$. Then
\begin{eqnarray}
Z^{M_\sigma}_{\xi_\sigma, \alpha_\sigma} =  \bar{Z}^{M_\sigma}_{\xi_\sigma, \alpha_\sigma}  / \sqrt{\lambda_{M_\sigma, \xi_\sigma}}
\end{eqnarray}
In the orthonormal basis the eigenvalue equation of the Hamiltonian for a given angular momentum z-component M can be written by the matrix equation
\begin{eqnarray}
( \textbf{H}^M - E_M \textbf{I} ) \boldsymbol{\psi}^M = 0,
\end{eqnarray}
where $\textbf{I}$ is the unit matrix and $\boldsymbol{\psi}^M$ is a column vector. The matrix elements of the matrix $\textbf{H}^M$ are given by 
\begin{align}
&\langle \xi_\nu^\prime M_\nu^\prime, \xi_\pi^\prime M_\pi^\prime; M| H |\xi_\nu M_\nu, \xi_\pi M_\pi; M \rangle \nonumber\\
=& \Big[\delta_{\xi_\pi^\prime, \xi_\pi }   \langle \xi_\nu^\prime M^\prime_\nu | \big( H_0(\nu) + H(\nu)\big)| \xi_\nu M_\nu \rangle +(\pi \leftrightarrow \nu)  \Big] \delta_{M_\nu^\prime,M_\nu}\delta_{M_\pi^\prime,M_\pi}\nonumber \\
&+ \langle \xi_\nu^\prime M_\nu^\prime, \xi_\pi^\prime M_\pi^\prime; M|  H_{\pi\nu} |\xi_\nu M_\nu, \xi_\pi M_\pi; M \rangle
\end{align}
Suppose that $\boldsymbol{\psi}^M_\beta=\psi^M_\beta(\xi_\nu M_\nu,\xi_\pi M_\pi)$ is the eigenvector of the matrix $\textbf{H}^M$ corresponding to the eigenvalue $E_{M_\beta}$, $\beta=1,~2,\dots$. Then the eigenfunction of the Hamiltonian with definite M is 
\begin{align}
|\beta,M\rangle = \sum\limits_{\xi_\nu M_\nu \xi_\pi M_\pi} \psi^M_\beta(\xi_\nu M_\nu,\xi_\pi M_\pi) | \xi_\nu M_\nu,\xi_\pi M_\pi \rangle
\end{align}
By using Eq.(\ref{eq_orthonormal}), eigenfunction can be expressed in terms of the non-orthonormal basis,
\begin{align}
|\beta,M\rangle &=   \sum\limits_{\xi_\nu M_\nu \xi_\pi M_\pi} \psi^M_\beta(\alpha_\nu \xi_\nu M_\nu, \alpha_\nu\xi_\pi M_\pi) Z^{M_\nu}_{\xi_\nu, \alpha_\nu} Z^{M_\pi}_{\xi_\pi, \alpha_\pi}   | \alpha_\nu M_\nu,\alpha_\pi M_\pi \rangle \nonumber\\
&= \sum\limits_{\alpha_\nu M_\nu \alpha_\pi M_\pi}  X^M_\beta(\alpha_\nu M_\nu \alpha_\pi M_\pi)   | \alpha_\nu M_\nu,\alpha_\pi M_\pi \rangle
\end{align}

\section{NPSM in M-scheme with isospin symmetry}\label{with_isospin}
In this section, the NPSM with isospin symmetry in m-scheme is presented. The uncoupled commutators, and matrix elements of one-body operators and two body operators are presented. The odd system and even system are treated on the same foot.

\subsection{Uncoupled commutators for nucleons pairs}
We begin by introducing the notation for system with isospin. The creation operator of a nucleon in a state with total angular momentum $j$, projection of angular momentum $m$, the isospin $t$ and projection of isospin $\tau$ is designated as $C^{jt\dagger}_{m\tau}$. Now we can build non-collective pair creation operator,
\begin{align}
A^{JT\dagger}_{m\tau}(ab) &= (C^{a\dagger}\times C^{b\dagger})^{JT}_{m\tau}\\
&= \sum\limits_{\substack{
m_a m_b\\
\tau_a \tau_b}}C^{Jm}_{j_a m_a,j_b m_b}C^{T\tau}_{t_a \tau_a,t_b \tau_b} C^{j_a t_a\dagger}_{m_a \tau_a}\times C^{j_b t_b\dagger}_{m_b \tau_b}  \nonumber
\end{align}
where $a$ represents $j_a$ and $t_a$, and $C^{Jm}_{j_a m_a,j_b m_b}$ is the Clebsch-Gordan coefficient. The single annihilation operator can be introduced as $C^{jt}_{m\tau} \equiv (C^{jt\dagger}_{m\tau})^\dagger $. The time-reversed form of single annihilation operator is
\begin{align}
\tilde C^{jt}_{m\tau} = (-)^{j+t-m-\tau}C^{jt}_{-m-\tau}
\end{align}
The commutator between single nucleon operator in coupled form is given by
\begin{align}
(\tilde{C}^a,C^{b\dagger})^{JT}_{m\tau} = \delta_{J0}\delta_{T0}\delta_{j_a j_b}\delta_{t_a t_b} \sqrt{2} \hat{j}_a
\end{align}

The non-collective pair annihilation operator is given by $A^{JT}_{m\tau}(ab) = ( A^{JT\dagger}_{m\tau}(ab) )^\dagger$, and the corresponding time-reversed form of pair annihilation operator is designated as,
\begin{align}
\tilde A^{JT}_{m\tau}(ab) &= (-)^{J+T-m-\tau}A^{JT}_{-m-\tau}(ab) \\
&=-(\tilde C^{a}\times \tilde C^{b})^{JT}_{m\tau} \nonumber
\end{align}

The collective pair creation, annihilation and time-reserved form operators are
\begin{align}
A^{JT\dagger}_{m\tau} &= \sum\limits_{ab}y(abJT)A^{JT\dagger}_{m\tau}(ab) \\
A^{JT}_{m\tau} &= \sum\limits_{ab}y(abJT) A^{JT}_{m\tau}(ab) \nonumber\\
\tilde A^{JT}_{m\tau} &= \sum\limits_{ab}y(abJT)\tilde A^{JT}_{m\tau}(ab) \nonumber
\end{align}
where y(abJT) is the pair structure coefficients, and has the symmetry as
\begin{align}
y( abJT) = (-)^{J+T-j_a-j_b} y(baJT)
\end{align}

The collective multipole operator is defined by,
\begin{align}
Q^{JT}_{m\tau} &= \sum\limits_{ab}q(abJT)Q^{JT}_{m\tau}(ab) \\
&=\sum\limits_{ab}q(abJT) ( C^{a\dagger} \times \tilde C^b )^{JT}_{m\tau}\nonumber
\end{align}

The uncoupled commutators for collective pairs can also obtained through coupled commutators. The coupled commutators in the NPSM with isospin symmetry have already been carried out in Ref.\cite{PhysRevC.87.044310}. The uncoupled commutator between two collective pair can be given by
\begin{align}
\label{eq_isospin_pairpair}
[A^{rt_1}_{\mu \tau_1},A^{s t_2\dagger}_{\nu \tau_2}] &= (-)^{r+t_1+\mu+\tau_1}\sum\limits_{\substack{
J m    \\
T\tau}} C^{Jm}_{r -\mu, s\nu}C^{T\tau}_{t_1 -\tau_1,t_2 \tau_2}[A^{rt_1},A^{s t_2 \dagger}]^{JT}_{m\tau}\\
&=2\delta_{rs}\delta_{t_1 t_2}\delta_{\mu \nu}\delta_{\tau_1 \tau_2} \sum\limits_{ab} y(abrt_1)y(abst_2)  \nonumber \\
&~~~~+  4(-)^{r+t_1+\mu+\tau_1}\sum\limits_{\substack{
J m\\
T\tau}} C^{Jm}_{r -\mu, s\nu} C^{T\tau}_{t_1 -\tau_1,t_2 \tau_2}  P^{JT}_{m\tau} \nonumber
\end{align}
where the summation run over projection m and $\tau$ are redundant, since it is a constant value $m = \nu-\mu$ and $\tau = \tau_2 - \tau_1$, and $P^{JT}_{m\tau}$ is a new one-body operator,
\begin{align}\label{eq_isospin_newonebody}
& P^{JT}_{m\tau} = \sum_{da}\hat r \hat s \hat{t}_1 \hat{t}_2    \left\{
   \begin{aligned}
&\frac{1}{2} &\frac{1}{2}~~~&t_1 \\
&t_2 &T ~~~&\frac{1}{2} \end{aligned}
    \right  \}
&\times  \sum_b y(abrt_1)y(bdst_2) \left\{
   \begin{aligned}
&j_a &j_b~~~&r_1 \\
&s &J ~~~&j_d \end{aligned}
    \right  \}  ( C^{d\dagger} \times \tilde{C}^a )^{JT}_{m\tau}
\end{align}

The uncouple commutator for collective pair and multipole operator is given by,
\begin{align}\label{eq_isospin_paironebody}
&[A^{rt_1}_{\mu \tau_1},Q^{l t_2}_{\sigma \tau_2}]\\
=&-\sum\limits_{\substack{
J T\\
m\tau}}(-)^{J+T+r_1+t_1-\sigma-\tau_2}C^{Jm}_{r_1 -\mu,l\sigma}C^{T\tau}_{t_1 -\tau_1,t_2\tau_2} \mathbb{A}^{JT}_{-m-\tau} \nonumber
\end{align}
where $\mathbb{A}^{JT}_{-m-\tau}$ is a new pair, which is
\begin{align}
&\mathbb{A}^{JT}_{-m-\tau} = \sum\limits_{da}y^\prime(daJT) \big( C^d \times C^a \big)^{JT}_{-m-\tau}  \\
&y^\prime(daJT) = z(daJT) +(-)^{J+T-j_a-j_d}z(adJT)  \nonumber\\
&z(daJT)= \hat{r}_1 \hat{l} \hat{t}_1 \hat{t}_2   \left\{
   \begin{aligned}
&\frac{1}{2} &\frac{1}{2}~~~&t_1 \\
&t_2 &T ~~~&\frac{1}{2} \end{aligned}
    \right  \} \sum\limits_{b}y(abr_1t_1)q(bdlt_2)\left\{
   \begin{aligned}
&j_a &j_b~~~&r_1 \\
&j &J ~~~&j_d \end{aligned}
    \right  \} \nonumber
\end{align}

By using Eq.(\ref{eq_isospin_pairpair}) and (\ref{eq_isospin_paironebody}) the uncoupled double commutator can be obtained,
 \begin{align}
\label{eq_isospin_pairdouble}
&\bigg[ A^{r_it_i}_{\mu_i \tau_i} ,\big[A^{r_kt_k}_{\mu_k \tau_k},A^{s t\dagger}_{\nu \eta}\big] \bigg] \\
=& -4 \sum\limits_{\substack{
~l m^\prime    \\
t^\prime \tau^\prime}} \sum\limits_{\substack{
J m    \\
T\tau}}(-)^{J+T+r_k+r_i+t_k+t_i-\nu-\eta} C^{l m^\prime}_{r_k -\mu_k, s \nu}C^{t^\prime\tau^\prime}_{t_k -\tau_k,t \eta} C^{J m}_{r_i -\mu_i, l m^\prime} C^{T\tau}_{t_i -\tau_i,t^\prime \tau^\prime}   \mathbb{B}^{JT}_{-m-\tau}\nonumber
\end{align}
where the summation runs over projection $m^\prime$, $\tau^\prime$, $m$ and $\tau$ is redundant, and $\mathbb{B}^{JT}_{-m-\tau}$ is a new pair,
\begin{align}
&\mathbb{B}^{JT}_{-m-\tau} = \sum\limits_{aa^\prime}y^\prime(aa^\prime JT) \big( C^a \times C^{a^\prime} \big)^{JT}_{-m-\tau}  \\
&y^\prime(aa^\prime JT) = z(aa^\prime JT) +(-)^{J+T-j_a-j_{a^\prime}}z(aa^\prime JT)  \nonumber\\
&z(aa^\prime JT)= \hat{r}_k \hat{r}_i \hat{s} \hat{l} \hat{t}_k  \hat{t}_i \hat{t} \hat{t}^\prime  \left\{
   \begin{aligned}
&\frac{1}{2} &\frac{1}{2}~~~&t_k \\
&t &t^\prime ~~~&\frac{1}{2} \end{aligned}
 \right  \} \left\{
   \begin{aligned}
&\frac{1}{2} &\frac{1}{2}~~~&t_i \\
&t^\prime &T ~~~&\frac{1}{2} \end{aligned}
 \right  \}   \sum\limits_{b b^\prime} y(a b^\prime r_k t_k) y(a^\prime b r_i t_i) y(b^\prime b s t) \nonumber \\
    & ~~~~~~~~~~~~~~~\times  \left\{
   \begin{aligned}
&j_a &j_{b^\prime}~~~&r_k \\
&s &l ~~~&j_b \end{aligned}
    \right  \}  \left\{
   \begin{aligned}
&j_{a^\prime} &j_{b}~~~&r_i \\
&l &J ~~~&j_a \end{aligned}
    \right  \} \nonumber
\end{align}

We can also obtain the uncoupled commutators between the single nucleon operator and multipole operator,
\begin{align} \label{eq_isospin_singleQ}
&[ C^{j_0 t_0}_{m_0 \tau_0}, Q^{j_2 t_2}_{\sigma \tau_2}] \\
=&  (-)^{j_2 -\sigma } (-)^{ t_2 - \tau_2 }   \sum\limits_{\substack{
J m    \\
T\tau}}    C^{Jm}_{j_0 -m_0, j_2 \sigma}   C^{T\tau}_{t_0 -\tau_0, t_2 \tau_2}     \frac{ \hat{j}_2  \hat{t}_2}{ \hat J \hat{ T} } q\big(  (j_0t_0) (JT) j_2t_2 \big) C^{JT}_{-m-\tau}   \nonumber
\end{align}

Base on  Eq.(\ref{eq_isospin_singleQ}), the uncoupled commutator between single nucleon operator and multipole-multipole interaction operator is given as,
 \begin{align} \label{eq_isospin_singleQQ}
& \sum\limits_{\sigma \tau_2} (-)^{\sigma + \tau_2} \bigg[  \big[ C^{j_0 t_0}_{m_0 \tau_0}, Q^{j_2 t_2}_{\sigma \tau_2} \big], Q^{j_2 t_2}_{-\sigma -\tau_2} \bigg] \\
=&  \sum\limits_{lt}   (-)^{j_0 -l} (-)^{t_0 - t } \frac{ ( 2j_2 + 1 )( 2t_2 +1 ) }{ ( 2j_0 + 1 )( 2t_0 +1 ) } q\big(  (j_0t_0) (lt) j_2t_2 \big) q\big( (lt) (j_0t_0)  j_2t_2 \big) C^{j_0 t_0}_{m_0\tau_0}   \nonumber
\end{align}

The uncoupled double commutator for single nucleon operator can also be obtained,
 \begin{align} \label{eq_isospin_singledouble}
& \bigg[  \big[ C^{j_0 t_0}_{m_0 \tau_0}, A^{r_k t_k}_{m_k \tau_k} \big], A^{s t}_{\nu \eta} \bigg] \\
=& 4\sum\limits_{\substack{
J m    \\
T\tau}}\sum\limits_{\substack{
l m^\prime    \\
t^\prime\tau^\prime}} (-)^{l +r_k -\nu} (-)^{t^\prime +t_k -\eta }C^{lm^\prime}_{r_k-m_k,s\nu} C^{t^\prime \tau^\prime}_{ t_k -\tau_k, t \eta  } C^{Jm}_{j_0 -m_0, l m^\prime } C^{T\tau}_{t_0 -\tau_0,t^\prime \tau^\prime} \frac{ \hat{r}_k \hat s \hat l }{ \hat J } \frac{ \hat{t}_k \hat t \hat{t}^\prime }{ \hat{T} }  \nonumber \\
 & \times  \left\{
   \begin{aligned}
&\frac{1}{2} &\frac{1}{2}~~~&t_k \\
&t &t^\prime ~~~&\frac{1}{2} \end{aligned}
 \right  \}  \sum\limits_{b}  y\big( (JT) b r_k t_k  \big )  y\big( b (j_0t_0)  s t  \big )  \left\{
   \begin{aligned}
&r_k &s~~~&l \\
&j_0 &J ~~~&j_b \end{aligned}
 \right  \}  C^{JT}_{-m-\tau} \nonumber
\end{align}
where the summation over projection $m$, $m^\prime$, $\tau$ and $\tau^\prime$ is redundant.

\subsection{commutators  in M-scheme}
In this section, the odd system with $2N+1$ nucleons and even system with $2N$ nucleons are treated on the same foot. The configuration space of the NPSM are constructed by collective pairs. And the creation operator  with specific total angular momentum  projection $M$ and total isospin projection $\tau$ are designed as
\begin{align}\label{eq_isospin_basis_convention}
&A^\dagger(\mathbbm{r}_0 , \dots,\mathbbm{r}_N )_{M,\tau }
\equiv A^\dagger(r_0 m_0 t_0 \tau_0, \dots,r_N m_N t_0 \tau_0 )_{M,\tau } \\
=& A^{r_0t_0\dagger}_{m_0 \tau_0}\cdot A^{r_1t_1 \dagger}_{m_1 \tau_1}\dots A^{r_N t_N \dagger}_{m_N \tau_N},\\
&~~~~~M = \sum\limits_{i=0}^{N}m_i,  ~~~~~~ \tau = \sum\limits_{i=0}^{N}\tau_i \nonumber
\end{align}
where the $\mathbbm{r}_0$ express all quantum number of this pair, and the convention  used for the operator $A^{r_0 t_0\dagger}_{m_0 \tau_0}$ is given as following,
\begin{align}
A^{r_0 t_0\dagger}_{m_0 \tau_0} = \left\{  \begin{aligned}
&1     &~~~\text{for even system,  }&m_0 \equiv 0,~\tau_0 \equiv 0 \\
&C^{r_0\dagger}_{m_0}  &~~~\text{for odd system,  } &r_0 \equiv j  \end{aligned} \right.
\end{align}
The annihilation operator is designated by
\begin{align}
&A(\mathbbm{r}_0 , \dots,\mathbbm{r}_N )_{M,\tau }
\equiv A(r_0 m_0 t_0 \tau_0, \dots,r_N m_N t_0 \tau_0 )_{M,\tau } \\
=& A^{r_0t_0}_{m_0 \tau_0}\cdot A^{r_1t_1 }_{m_1 \tau_1}\dots A^{r_N t_N }_{m_N \tau_N},\\
&~~~~~M = \sum\limits_{i=0}^{N}m_i,  ~~~~~~ \tau = \sum\limits_{i=0}^{N}\tau_i \nonumber
\end{align}
where the convention of $A^{r_0}_{m_0}$ is similar to that of Eq.(\ref{eq_isospin_basis_convention}). The time-reversed form of $A(\mathbbm{r}_0 , \dots,\mathbbm{r}_N )_{M,\tau }$ is redundant in this model.

Then we can obtain the commutator between the annihilation operator$A(\mathbbm{r}_0 , \dots,\mathbbm{r}_N )_{M,\tau } $
\begin{align}\label{eq_isospin_basisandpair}
&[A(\mathbbm{r}_0,\mathbbm{r}_1,\dots,\mathbbm{r}_N)_{M,\tau}~,A^{st \dagger}_{\nu \eta}]   \\
=&\sum\limits_{k=1}^{N} \bigg[ \varphi \delta_{r_k,s}\delta_{m_k,\nu} \delta_{t_k,t} \delta_{t_k,\eta}   A(\mathbbm{r}_0,\dots,\mathbbm{r}_{k-1} , \mathbbm{r}_{k+1},\dots, \mathbbm{r}_N)_{M-\nu,\tau-\eta}  \nonumber \\
+&\sum\limits_{i=k-1}^{0~or~1}\sum\limits_{~r^\prime_i m^\prime_i} \sum\limits_{~t^\prime_i \tau^\prime_i} A(\mathbbm{r}_0 ,\dots, \boldsymbol{\mathbbm{r}}_{i}^\prime ,\dots ,\mathbbm{r}_{k-1},\mathbbm{r}_{k+1},\dots, \mathbbm{r}_N)_{M-\nu,\tau-\eta} \nonumber\\
+& \sum\limits_{l m^\prime}\sum\limits_{t^\prime \tau^\prime} P^{lt^\prime}_{m^\prime \tau^\prime} \times A(\mathbbm{r}_0,\dots ,\mathbbm{r}_{k-1} ,\mathbbm{r}_{k+1} ,\dots, \mathbbm{r}_N)_{M-m_k,\tau-\tau_k}  \bigg]\nonumber
\end{align}
where $\varphi=2 \sum\limits_{ab}y(a b r_k t_k)y(a b st)$, summation  over $i$ is from $k-1$ to $0$ or $1$ corresponding to odd  system and even  system respectively, and $\boldsymbol{\mathbbm{r}}_{i}^\prime$ is a new collective pair $A^{r^\prime_i t^\prime_i}_{m^\prime_i \tau^\prime_i}$ ($i\ne0$) or a single nucleon operator $C^{r^\prime_0 t^\prime_0}_{m^\prime_0 \tau^\prime_0}$ ($i=0$), and can be obtained from the double commutator by
\begin{align}
A^{r^\prime_i t^\prime_i}_{m^\prime_i \tau^\prime_i} = \bigg[ A^{r_i t_i}_{m_i \tau_i} ,\big[A^{r_k t_k}_{m_k \tau_k} ,A^{s t\dagger}_{\nu \eta} \big]\bigg]
\end{align}
the explicit form of the new pair has been given in Eq.(\ref{eq_isospin_pairdouble}) and (\ref{eq_isospin_singledouble}),  $P^{lt^\prime}_{m^\prime \tau^\prime}$ has been given in Eq.(\ref{eq_isospin_newonebody}).

By using Eq.(\ref{eq_isospin_basisandpair}), the commutation relation between general pairing interaction and $A(\mathbbm{r}_0,\dots,\mathbbm{r}_N)_{M,\tau}$ is obtained by
\begin{align}\label{eq_isospin_basisparing}
&[A(\mathbbm{r}_0,\dots,\mathbbm{r}_N)_{M,\tau}~,A^{st \dagger}\cdot A^{st}]\\
=&\sum\limits_{k=1}^{N} \bigg[ \varphi  \delta_{r_k,s}\delta_{m_k,m} \delta_{t_k,t} \delta_{t_k,\eta}  A(\mathbbm{r}_0,\dots,\mathbbm{s},\dots, \mathbbm{r}_N)_{M,\tau}  \nonumber \\
+&\sum\limits_{i=k-1}^{0~or~1}\sum\limits_{~r^{\prime}_i m^{\prime}_i m}  \sum\limits_{~t^{\prime}_i \tau^{\prime}_i \eta}  A(\mathbbm{r}_0,\dots, \boldsymbol{\mathbbm{r}}_{i}^\prime ,\dots ,\mathbbm{r}_{k-1},\mathbbm{s},\mathbbm{r}_{k+1} ,\dots, \mathbbm{r}_N)_{M,\tau}  \nonumber\\
+&\sum\limits_{l m^\prime m}\sum\limits_{t^\prime \tau^\prime \eta} P^{lt^\prime}_{m^\prime \tau^\prime}  \times A(\mathbbm{r}_0,\dots ,\mathbbm{r}_{k-1}, \mathbbm{s},\mathbbm{r}_{k+1} ,\dots, \mathbbm{r}_N )_{M-m_k+m, \tau-\tau_k+\eta} \bigg] \nonumber
\end{align}

The commutator between one body operator and $A(\mathbbm{r}_0,\dots,\mathbbm{r}_N)_{M,\tau}$ is obtained, which is 
\begin{align}\label{eq_isospin_basis_onebody}
&[A(\mathbbm{r}_0,\dots,\mathbbm{r}_N)_{M},Q^{jt}_{\sigma \eta}]  \\
=&\sum\limits_{k=N}^{0~or~1} \sum\limits_{r^\prime_k m^\prime_k} \sum\limits_{t^\prime_k \tau^\prime_k} A(\mathbbm{r}_0, \mathbbm{r}_1,\dots,\boldsymbol{\mathbbm{r}}^\prime_k,\dots,\mathbbm{r}_N)_{M-\sigma,\tau -\eta} \nonumber
\end{align}
where summation  over $k$ is from $N$ to $0$ or $1$ corresponding to odd  system or even  system respectively. $\boldsymbol{\mathbbm{r}}^\prime_k$ represents a new collective pair $A^{r^\prime_k t^\prime_k}_{m^\prime_k \tau^\prime_k}$ ($k\ne0$) or a single nucleon operator $C^{r^\prime_0 t^\prime_0}_{m^\prime_0 \tau^\prime_0}$ ($k=0$),
\begin{align}
A^{r^\prime_k t^\prime_k}_{m^\prime_k \tau^\prime_k} = [A^{r_k t_k}_{m_k \tau_k}, Q^{jt}_{\sigma \eta}]
\end{align}
The explicit form of the new pair has been given in Eq.(\ref{eq_isospin_paironebody}) and (\ref{eq_isospin_singleQ}).

The commutator between the general multipole-multipole operator and $A(\mathbbm{r}_0,\dots,\mathbbm{r}_N)_{M,\tau}$ is obtained by
\begin{align}\label{eq_isospin_basis_m-m}
&[A(\mathbbm{r}_0,\dots,\mathbbm{r}_N)_{M,\tau} ~, \sum\limits_{\sigma \eta} (-)^{\sigma + \eta} Q^{jt}_{\sigma \eta}  Q^{jt}_{-\sigma -\eta} ] \\
=&\sum\limits_{k=N}^{0~or~1}\bigg[ A(\mathbbm{r}_0 , \mathbbm{r}_1 ,\dots,(\boldsymbol{\mathbbm{r}}_k)_B,\dots,\mathbbm{r}_N)_{M}\nonumber\\
+&  \sum\limits_{i=k-1}^{0~or~1}  \sum\limits_{\substack{
r_i^\prime m_i^\prime\\
r_k^\prime m_k^\prime}} \sum\limits_{\substack{
t_i^\prime \tau_i^\prime\\
t_k^\prime \tau_k^\prime}}   \sum\limits_{\sigma \eta} (-)^{\sigma + \eta}    A(\mathbbm{r}_0, \mathbbm{r}_1 ,\dots,\boldsymbol{\mathbbm{r}}^\prime_i ,\dots,\boldsymbol{\mathbbm{r}}^\prime_k,\dots, \mathbbm{r}_N )_{M,\tau}  \nonumber\\
+&  \sum\limits_{
r_k^\prime m_k^\prime} \sum\limits_{
t_k^\prime \tau_k^\prime} \sum\limits_{\sigma\tau}  (-)^{\sigma+\eta} Q^{jt}_{-\sigma-\tau} \times A(\mathbbm{r}_0, \mathbbm{r}_1,\dots,\boldsymbol{\mathbbm{r}}^\prime_k ,\dots,\mathbbm{r}_N)_{M-\sigma,\tau-\eta} \bigg] \nonumber
\end{align}
where summation range over $k (i)$ is from $N (k-1)$ to $0$ or $1$ corresponding to odd  system or even  system respectively. $\boldsymbol{\mathbbm{r}}^\prime_k  (\boldsymbol{\mathbbm{r}}^\prime_i)$  represents a new collective pair ($k\ne0$) or a single nucleon ($k=0$),
\begin{align}
A^{r^\prime_k t^\prime_k }_{m^\prime_k \tau^\prime_k} = [A^{r_k  t_k}_{m_k \tau_k}, Q^{jt}_{\sigma\eta}] \nonumber \\
A^{r^\prime_i t^\prime_i}_{m^\prime_i \tau^\prime_i} = [A^{r_i t_i}_{m_i \tau_i}, Q^{jt}_{-\sigma -\eta}]
\end{align}
The explicit form of the new pair has been given in Eq.(\ref{eq_isospin_paironebody}) and (\ref{eq_isospin_singleQ}). $(\boldsymbol{\mathbbm{r}}_k)_B$ denotes a new collective pair ($k\ne0$) or a single nucleon ($k=0$), and can be obtained by uncoupled double commutator,
\begin{align}
&\big(A^{r_k t_k }_{m_k \tau_k}\big)_B =\sum\limits_{\sigma}(-)^\sigma  \bigg [ \big[A^{r}_{m},  Q^t_\sigma \big], Q^t_{-\sigma} \bigg]     \label{}  \\
&\big(A^{r_k t_k }_{m_k \tau_k}\big)_B =\bigg [C^j_{m_0}, \big[ A^{r_k}_{m_k},A^{s\dagger}_{m}  \big]\bigg]  \nonumber
\end{align}
The explicit results have been carried out in Eq.(\ref{eq_isospin_pairdouble}) and (\ref{eq_isospin_singleQQ}).

\subsection{Matrix elements of overlap and interactions}
A $N$-pair or $N$-pair plus one single nucleon state in M-scheme is designated as
\begin{align}
|\alpha,M\tau> &= |\mathbbm{r}_0,\mathbbm{r}_1,\dots, \mathbbm{r}_N;M,tau\rangle \\
&\equiv A^\dagger(\mathbbm{r}_0,\mathbbm{r}_1,\dots, \mathbbm{r}_N)|0\rangle \nonumber
\end{align}
where $\mathbbm{r}_k$ represents a pair with angular momentum $r_k$, angular momentum  projection $m_k$, isospin $t_k$ and isospin  projection $\tau_k$, $\mathbbm{r}_0$ denote $1$ in even  system or a single nucleon in odd system, $\alpha$ denotes the additional quantum numbers.
\begin{align}
\alpha =(r_0m_0 t_0 \tau_0,\dots,r_N,m_N t_N \tau_N)
\end{align}
It is interesting to note that $\alpha$ is redundant, which contain $4N$ quantum numbers and have already been included in the information of total projection M and $\tau$.

The overlap matrix element is a key quantity, since the one and two-body interaction matrix elements can be expressed as summation of the overlaps. From Eq.(\ref{eq_isospin_basisandpair}),  the overlap can be obtained by
\begin{align}\label{eq_isospin_overlap}
&  \langle 0|A_{M_l \tau_l}A^{\dagger}_{M_r \tau_r}|0\rangle  =\langle \mathbbm{r}_0,\dots,\mathbbm{r}_N;M_l \tau_l | \mathbbm{s}_0,\dots,\mathbbm{s}_N;M_r \tau_r \rangle  \\
\equiv & \langle r_0\mu_0t_0 \tau_0  ,\dots, r_N\mu_N t_N \tau_N  ;M_l \tau_l|s_0\nu_0  h_0 \eta_0,\dots, s_N\nu_N h_0 \eta_0;M_r \tau_r \rangle \nonumber \\
=&\sum\limits_{k=1}^{N} \bigg[ 2 \sum\limits_{ab}y(a b r_k t_k)y(a b s_Nh_N) \delta_{r_k,s_N}\delta_{\mu_k,\nu_N} \delta_{t_k,h_N}\delta_{\tau_k,\eta_N} \nonumber \\
&\times \langle \mathbbm{r}_0\dots \mathbbm{r}_{k-1},\mathbbm{r}_{k+1} \dots \mathbbm{r}_N ;M_l-\nu_N, \tau_l - \eta_N|\mathbbm{s}_0\dots \mathbbm{s}_{N-1};M_r-\nu_N, \tau_r - \eta_N \rangle  \nonumber \\
&+\sum\limits_{i=k-1}^{0~or~1}\sum\limits_{~r^\prime_i \mu^\prime_i}\sum\limits_{~t^\prime_i \tau^\prime_i}
\nonumber \\ &\times   \langle \mathbbm{r}_0 \dots~  \boldsymbol{\mathbbm{r}}^\prime_{i}  \dots \mathbbm{r}_{k-1}, \mathbbm{r}_{k+1} \dots \mathbbm{r}_N ;M_1-\nu_N , \tau_l -\eta_N|\mathbbm{s}_0\dots \mathbbm{s}_{N-1};M_r-\nu_N, \tau_r - \eta_N \rangle \bigg] \nonumber
\end{align}
where $r_k$ and $s_k$ denote the angular moment of the $k$-th pair, and $\mu_k$ and $\nu_k$ denote its  projection of angular moment, $t_k$ and $h_k$ represent the isospin, and $\tau_k$ and $\eta_k$ stand for its projection.
Although the overlap matrix elements formula in M-scheme is still written in a recursive way, it does not need to recouple the  angular momentum and  isospin.
It is easy to show the total angular momentum projection $M_l$ should be equal to $M_r$, and  total projection of isospin $\tau_l$ should be equal to $\tau_r$.
Notice that the summation over the projections of the new pair, $\mu^\prime_i$ and $\tau^\prime_i$, are redundant, which should be a constant value $\mu_i+\mu_k-\nu_N$ and $\tau_i+\tau_k-\eta_N$, respectively, since projection is a scalar. The  overlap  for one pair is given by using Eq.(\ref{eq_isospin_overlap}), which is
\begin{align}
\langle \mathbbm{r}_1  | \mathbbm{s}_1 \rangle
=2\delta_{r_1,s_1} \delta_{\mu_1,\nu_1}\delta_{t_1,h_1} \delta_{\tau_1,\eta_1} \sum\limits_{ab} y(abr_1t_1)y(abs_1h_1)
\end{align}
Since there is only one pair, the overlap formula is equivalent to the formulae for the case of $N = 1$ in J-scheme. The overlap for the one pair plus one single nucleon is given as following,
\begin{align}
&\langle \mathbbm{r}_0, \mathbbm{r}_1 ;M\tau | \mathbbm{s}_0 ,\mathbbm{s}_1 ;M\tau\rangle  \\
=&2\delta_{r_1,s_1} \delta_{\mu_1,\nu_1}\delta_{t_1,h_1} \delta_{\tau_1,\eta_1} \sum\limits_{ab} y(abr_1t_1)y(abs_1h_1)  \nonumber \\
& +4\sum\limits_{\substack{
l m    \\
t^\prime\tau^\prime}} (-)^{l +r_1 -\nu_1} (-)^{t^\prime +t_1 -\eta_1 }C^{lm}_{r_1-\mu_1,s_1\nu_1} C^{t^\prime \tau^\prime}_{ t_1 -\tau_1, h_1 \eta_1  } C^{s_0 -\nu_0}_{r_0 -\mu_0, l m } C^{h_0-\eta_0}_{t_0 -\tau_0,t^\prime \tau^\prime} \frac{ \hat{r}_1 \hat {s_1} \hat l }{ \hat{s}_0 } \frac{ \hat{t}_1 \hat{h}_1 \hat{t}^\prime }{ \hat{h}_0 }  \nonumber \\
 & \times  \left\{
   \begin{aligned}
&\frac{1}{2} &\frac{1}{2}~~~&t_1 \\
&h_1 &t^\prime ~~~&\frac{1}{2} \end{aligned}
 \right  \}  \sum\limits_{b}  y\big( (s_0h_0) b r_1 t_1  \big )  y\big( b (r_0t_0)  s_1 h_1  \big )  \left\{
   \begin{aligned}
&r_1 &s_1~~~&l \\
&r_0 &s_0 ~~~&j_b \end{aligned}
 \right  \} \nonumber
\end{align}
where $\mathbbm{r}_0(\mathbbm{s}_0)$ denotes the single nucleon.

It is easy to show that the matrix elements of a pair creation operator $A^{s\dagger}_\nu$ between two states differing by one pair is equal to an overlap,
\begin{align}
&\langle \mathbbm{r}_0,\mathbbm{r}_1, \dots,\mathbbm{r}_N |A^{s h\dagger}_{\nu \eta} | \mathbbm{s}_0,\mathbbm{s}_1, \dots,\mathbbm{s}_{N-1} \rangle  \\
=& \langle \mathbbm{r}_0 ,\mathbbm{r}_1,\dots,r_N \mu_N| \mathbbm{s}_0 , \mathbbm{s}_1 ,\dots,\mathbbm{s}_{N-1}, \mathbbm{s}\rangle \nonumber
\end{align}

Using Eq.(\ref{eq_isospin_basisparing}), we can have the matrix elements of pairing interaction,
\begin{align}\label{eq_pairing_interaction_isospin}
 &\langle \mathbbm{r}_0,\mathbbm{r}_1,\dots \mathbbm{r}_N;M\tau|A^{st\dagger}\cdot A^{st}|\mathbbm{s}_0,\mathbbm{s}_1,\dots, \mathbbm{s}_N;M \tau \rangle \\
=&\sum\limits_{k=1}^{N} \bigg[ \varphi \delta_{r_k,s}\delta_{\mu_k,m} \delta_{t_k,t}\delta_{\tau_k,\eta}  \langle \mathbbm{r}_0 \dots~ \boldsymbol{\mathbbm{s}}\dots \mathbbm{r}_N;M \tau|\mathbbm{s}_0 \dots \mathbbm{s}_{N};M\tau\rangle  \nonumber \\
+&\sum\limits_{i=k-1}^{0~or~1}\sum\limits_{~r^\prime_i \mu^\prime_i m} \sum\limits_{~t^\prime_i \tau^\prime_i \eta} \langle \mathbbm{r}_0 \dots~ \boldsymbol{\mathbbm{r}}^\prime_i \dots~ \boldsymbol{\mathbbm{s}}  \dots \mathbbm{r}_N ;M \tau|\mathbbm{s}_0\dots \mathbbm{s}_{N};M \tau \rangle \bigg] \nonumber
\end{align}
where the summation over the projection $m$ and $\eta$ in the second term standing for the projection of pairing interacting $G_{st}\sum\limits_{m \eta}A^{st\dagger}_{m\eta} A^{st}_{m\eta}$, and the summation over the projection $\mu^\prime_i$ and $\tau^\prime_i$ of new pair is redundant, which should be constant values $\mu_i+\mu_k-m$ and $\tau_i+\tau_k-\eta$, respectively.

By using Eq.(\ref{eq_isospin_basis_onebody}), for the one-body operator matrix element we have
\begin{align}
& \langle \mathbbm{r}_0 \dots\mathbbm{r}_N  ;{M_1}\tau_1| Q^{jt}_{\sigma\eta} | \mathbbm{s}_0\dots \mathbbm{s}_N;M_2 \tau_2 \rangle  \\
=&\sum\limits_{k=N}^{0~or~1}\sum\limits_{r^\prime_k ,\mu^\prime_k} \sum\limits_{t^\prime_k ,\tau^\prime_k} \langle \mathbbm{r}_0 \dots~\boldsymbol{\mathbbm{r}}^\prime_k \dots \mathbbm{r}_N ;M_1-\sigma,~\tau_1 -\eta|\mathbbm{s}_0 \dots \mathbbm{s}_N ;M_2 \rangle \nonumber
\end{align}
where the summation over the projection $\mu^\prime_k$ and $\tau^\prime_k$ of new pair is redundant, which should be a constant value $\mu_k-\sigma$ and $\tau_k-\eta$, respectively.

 By using Eq.(\ref{eq_isospin_basis_m-m}), we can obtain the matrix elements of the multipole-multipole interaction between like nucleons,
\begin{align}\label{eq_QQ_interaction_isospin}
& \langle \mathbbm{r}_0 \dots, \mathbbm{r}_N  ;{M} \tau| Q^{jt} \cdot Q^{jt} | \mathbbm{s}_0\dots \mathbbm{s}_N ;M \tau  \rangle \\
=&\sum\limits_{k=N}^{0~or~1}\bigg[  \langle \mathbbm{r}_0\dots(\mathbbm{r}_k )_B\dots \mathbbm{r}_N ;M \tau |  \mathbbm{s}_0\dots \mathbbm{s}_N ;M \tau \rangle \nonumber   \\
+&  \sum\limits_{i=k-1}^{0~or~1}  \sum\limits_{\substack{
r_i^\prime m_i^\prime\\
r_k^\prime m_k^\prime}}   \sum\limits_{\substack{
t_i^\prime \tau_i^\prime\\
t_k^\prime \tau_k^\prime}}\sum\limits_{\sigma \eta} (-)^{\sigma +\eta} \langle \mathbbm{r}_0 \dots \mathbbm{r}^\prime_i \dots \mathbbm{r}^\prime_k \dots \mathbbm{r}_N ;M\tau |\mathbbm{s}_0 \dots \mathbbm{s}_N;M\tau \rangle  \bigg] \nonumber
\end{align}
where the summation over the projection $\mu^\prime_k~(\mu^\prime_i)$ and $\tau^\prime_k~(\tau^\prime_i)$ of the new pair is redundant, since it should be a constant values $\mu_k-\sigma~(\mu_k+\sigma)$ and $\tau_k-\eta~(\tau_k+\eta)$, respectively.

\subsection{Diagonalization of the Hamiltonian}
In this part, a general Hamiltonian are considered, which consist of the single-particle energy term $H_0$, and a residual interaction $V$, 
\begin{align}
&H =  H_0+ V   \\
&V=-\frac{1}{4}\sum\limits_{J,T= 0,1}\sum\limits_{a,b,c,d} [\mathbbm{N}_{ab}(JT) \mathbbm{N}_{cd}(JT)]^{-1} \hat J \hat T \langle ab;JT|V|cd;JT\rangle  \nonumber \\
&~~~~~~ \times \left[  [ C_a^\dagger  C_b^\dagger]^{TJ}[ \tilde C_c  \tilde C_d]^{TJ} \right] ^{00} \nonumber \\
&\mathbbm{N}_{ab}(JT) = \frac{ \sqrt{ 1- \delta_{ab}(-)^{J+T}} }{1+\delta_{ab}} \nonumber
\end{align}
where $\langle ab;JT|V|cd;JT\rangle$ are two-body matrix elements. The Hamiltonian can be diagonialized either in the non-orthonormal basis $|\alpha,  \tau, M\rangle$, or the orthonormal basis $|\xi, \tau, M\rangle$. The orthonormal basis $|\xi, \tau, M\rangle$  can be expanded in terms of the non-orthonormal basis  $|\alpha,  \tau, M\rangle$,
\begin{eqnarray}\label{eq_orthonormal_T}
|\xi, \tau, M\rangle = \sum_{\alpha} Z^{(\tau M)}_{\xi, \alpha}|\alpha, \tau, M\rangle
\end{eqnarray}
where the coefficients $Z^{(\tau M)}_{\xi, \alpha}$ are found in the following way. Suppose that $\bar {\textbf{Z}}^{(\tau M)}_{\xi}$ is an orthonormalized eigenvector of the overlap matrix $\Lambda^{(\tau M)}$,
\begin{eqnarray}
\Lambda^{(\tau M)}_{\alpha^\prime, \alpha} = \langle \alpha^\prime, \tau, M| \alpha, \tau, M\rangle
\end{eqnarray}
corresponding to the eigenvalue $\lambda_{\tau M, \xi}$. Then
\begin{eqnarray}
Z^{(\tau M)}_{\xi, \alpha} =  \bar{Z}^{(\tau M)}_{\xi, \alpha}   / \sqrt{\lambda_{\tau M, \xi}}
\end{eqnarray}
In the orthonormal basis the eigenvalue equation of the Hamiltonian for a given angular momentum z-component $M$ and isospin z-component $\tau$ can be written by the matrix equation
\begin{eqnarray}
( \textbf{H}^{(\tau M)} - E_{(\tau M)} \textbf{I} ) \boldsymbol{\psi}^{(\tau M)}  = 0,
\end{eqnarray}
where $\textbf{I}$ is the unit matrix and $\boldsymbol{\psi}^{(\tau M)}$ is a column vector. The matrix elements of the matrix $\textbf{H}^{(\tau M)}$ are given by 
\begin{align}
&\langle \xi^\prime, \tau, M| H |\xi, \tau, M \rangle
=\langle \xi^\prime, \tau, M| ( H_0+ V)|\xi, \tau, M \rangle
\end{align}
Suppose that $\boldsymbol{\psi}^{(\tau M)}_\beta=\psi^{(\tau M)}_\beta(\xi, \tau, M)$ is the eigenvector of the matrix $\textbf{H}^{(\tau M)}$ corresponding to the eigenvalue $E_{\tau M_\beta}$, $\beta=1,~2,\dots$. Then the eigenfunction of the Hamiltonian with definite $\tau$ and $M$ is 
\begin{align}
|\beta, \tau,M\rangle = \sum\limits_{\xi} \psi^{(\tau M)}_\beta(\xi, \tau, M) | \xi, \tau, M \rangle
\end{align}
By using Eq.(\ref{eq_orthonormal_T}), eigenfunction can be expressed in terms of the non-orthonormal basis,
\begin{align}
|\beta, \tau,M\rangle =   \sum\limits_{\xi, \alpha}   \psi^{(\tau M)}_\beta(\xi, \tau, M) Z^{(\tau M)}_{\xi, \alpha}|\alpha, \tau, M\rangle
\end{align}

\section{Discussion and summary}\label{summary}
The NPSM  is casted in the M-scheme  for both even and odd
 systems under the assumptions that all nucleon pairs are collective and for given angular momentum $r$
 there is only one collective pair $A_r^\dagger$. The cases with isospin symmetry and without isospin symmetry
 are all discussed.
 If there are more than one type of
collective pairs with given angular momentum $s$, we only need to introduce an additional
label to distinguish them. The NPSM in M-scheme can also be easily extended to include the
non-collective pairs.

It is known that all
the matrix elements in the NPSM in J-scheme can be expressed as the summation of overlaps between two $N$-pair basis.  The formulas to calculate the overlaps have to sum over all the intermediate angular momentum quantum numbers $J_i$ and  the quantum numbers of the new pairs.
Therefore, the $cpu$ time used in calculating the overlaps increases drastically with the number of nucleon pairs.

\begin{figure}
\includegraphics[width=10cm]{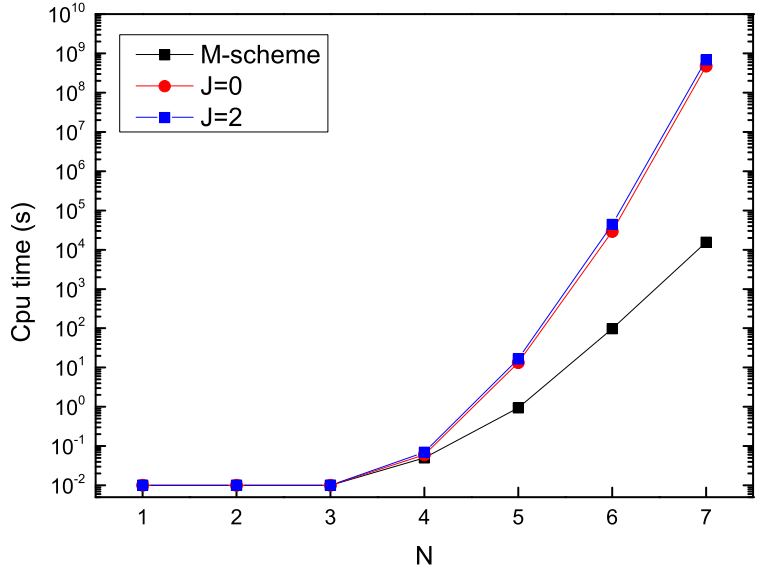}
\caption{The average $cpu$ time used in computing one overlap matrix element against the number of pairs are presented. The overlap in J-scheme are calculated in the subspace with total angular momentum $J=0$ and $J=2$. The overlap in M-scheme is for the states with $M=0$.}
\label{fig_overp}
\end{figure}

From our previous discussion, it is known that the formula in calculating matrix elements  in M-scheme do not contain the summation of the intermediate quantum number, 
 it is interesting to  see the validity of the NPSM in M-scheme. To this end
the average $cpu$ time used in computing the overlap versus number of pairs is presented in Fig.\ref{fig_overp}, in which  the average $cpu$ times of $16$ overlaps between the basis constructed by identical $D(J=2)$  collective pairs are presented. As example,
only the results for the systems without isospin symmetry are presented here.
Since the $cpu$ time is too small to be obtained for the systems with the number of the collective pairs smaller than 4, we set them all to be equal to $10^{-2}$ second per matrix element. And for the system with $N=7$,
the average $cpu$ time in calculating the overlap in J-scheme is too time-consuming to be obtained,
 its $cpu$ time is estimated through the recursive formula of the overlap and the average $cpu$ time of the overlap for the system with $N=6$.
Fig.\ref{fig_overp} shows  that the average $cpu$ time of overlap matrix element in M-scheme is indeed much smaller that those  in J-scheme. For example, the average $cpu$ time for the system with $N=6$ is about $10^2$ seconds in the M-scheme, while it is about $10^5$ seconds in the J-scheme.

\begin{table} [h]
\begin{tabular}{ c c | c c c c }\hline\hline
\multicolumn{2}{c|}{J-scheme} &  \multicolumn{4}{c}{M-scheme} \\ \hline
   ~~~~~J~~~~~ & Number of states &  ~~~~~M~~~~~ & Number of states &  ~~~~~M~~~~~ & Number of states   \\ \hline
   10	&	1   &  10	&	1  &	-10		&	1	\\
   8	&	2	 &	 9	&	1	&	-9		&	1	\\	
   7	&	1   &  8	&	3	&	-8		&	3  \\
   6	&	4	 &  7	&	4 	&	-7		&	4	\\
   5	&	2   &  6	&	8  &	-6		&	8	\\
   4	&	6	 &  5	&	10 &	-5		&	10 \\
   3	&	2   &  4	&	16 &	-4		&	16 \\
   2	&	7   &  3	&	18	&	-3  	&	18	\\
   0	&	5   &  2	& 	25	&	-2		&  25	\\
   		&		 &  1	&	25	&	-1  	&	25	\\
   		&		 &  0	&	30	&	  	&		\\  \hline \hline
\end{tabular}
\caption{Number of states for $5$ collective pairs in $SD$ subspace. The left column is for the case in J-scheme, while the right one is for the case in M-scheme.}
 \label{tab_Numberofstate}
\end{table}

To see the difference of the configuration space between M-scheme and J-scheme, the number of states for the system with $N=5$ collective $S(J=0)$ and $D(J=2)$ pairs are presented in Table.\ref{tab_Numberofstate}. It can be seen that the number of the states in J-scheme is much smaller than those in the M-scheme.

\begin{figure}
\includegraphics[width=10cm]{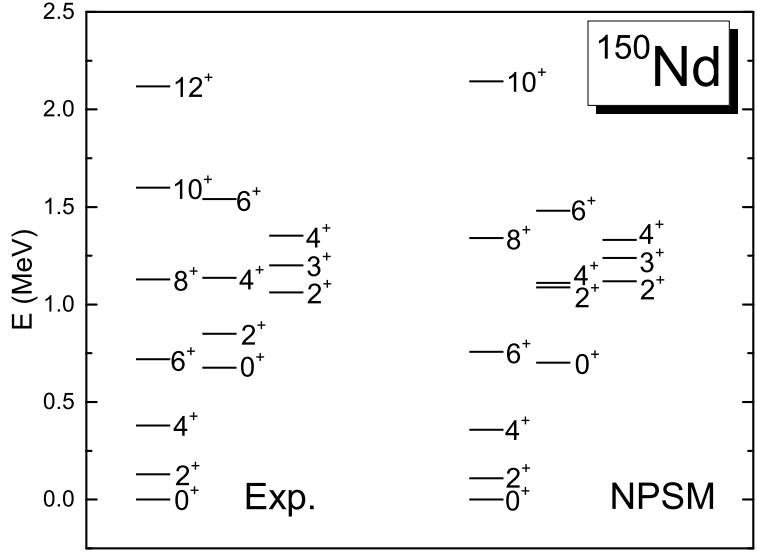}
\caption{The low-lying spectrum of $^{150}$Nd. The experimental data are obtained from Ref.\cite{nndc}.}
\label{fig_Nd}
\end{figure}

As a simple example, the pairing plus quadrupole-quadrupole interaction is employed to describe the low-lying spectrum of $^{150}$Nd in SD-pair subspace.
\begin{align}
\hat H &= \sum_{\sigma = \pi, \nu}\hat H_{\sigma} - \kappa Q_\pi^2 \cdot Q_\nu^2 \\
\hat H_{\sigma} &= H_{0 \sigma} - G_{0\sigma} A^{(0)\dagger} A^{(0)}  - G_{2\sigma} A^{(2)\dagger}  A^{(2)\dagger} - \kappa_\sigma Q^2 Q^2 \nonumber\\
H_{0 \sigma}  &= \sum_a \epsilon_a  C_a^\dagger C_a \nonumber\\
A^{(0)\dagger} &= \sum_a \frac{\hat a}{2} ( C_a^\dagger \times C_a^\dagger )^0 \nonumber\\
A^{(2)\dagger} &= \sum_{ab} q(ab2) ( C_a^\dagger \times C_b^\dagger )^2 \nonumber
\end{align}
where $H_{0 \sigma}$ is the single particle energy term, and $G_0$, $G_2$, $\kappa$ are the monopole pairing, the quadrupole pairing and the quadrupole-quadrupole interaction strength, respectively. And the exactly form of $q(ab2)$ have been given in Eq.(\ref{4.4b}). As an approximation, in this test the S-pair structure coefficient is chosen to be $y(aa0) = \hat j_a \frac{\nu_a}{\mu_a}$, where $\nu_a$ and $\mu_a$ are occupied and empty amplitude obtained by solving the BCS equation, while the D-pair structure coefficients are obtained from the commutator,
\begin{align}
A^{2\dagger} = D^\dagger = \frac{1}{2} [ Q^2, S^\dagger ]
\end{align}
The calculations are carried out in the model space spanned by $1f_{7/2}$,  $0h_{9/2}$,  $1f_{5/2}$,  $2p_{3/2}$,  $2p_{1/2}$,  $0i_{13/2}$ orbitals for neutrons and  $0g_{7/2}$,  $1d_{5/2}$, $1d_{3/2}$,  $2s_{1/2}$,  $0h_{11/2}$ orbitals for protons, taken above the closed core $^{132}$Sn.  The single-particle energies are adopted as $^{133}Sn$ and $^{133}$Sb experimental data\cite{nndc, PhysRevC.56.R16, UrbanNeutron}.

By the best fit to the experimental level energies in collective SD-pair subspace, the model parameters with $G_{0\pi}$ = 0.14 $MeV$, $G_{2\pi}$ = 0.07 $MeV / r_0^4$, $G_{0\nu}$ = 0.12 $MeV$, $G_{2\nu}$ = 0.04 $MeV / r_0^4$, $\kappa_\pi = \kappa_\nu$=0 and $\kappa$ = 0.17 $MeV / r_0^4$ are obtained.  As shown in the Fig.\ref{fig_Nd}, the low-lying spectum of $^{150}$Nd are reproduced by NPSM.  Effective charges of E2 transition operator are chosen as $e_\pi$ =2.3e for protons  and $e_\nu$ = 1.5e for neutrons.  Some B(E2) transition are listed in Table.\ref{tab_NdBE2}

\begin{table} [h]
\caption{$B(E2)$ calues for $^{150}$Nd.}
\begin{tabular}{ c c c c c c c }
\hline\hline
\multicolumn{7}{c}{ $B(E2; J_i \rightarrow J_f$ ~(eb)$^2$ }  \\
   $J_i \rightarrow J_f$	&	  & $B(E2)_{expt}$$^{a}$	&	  &	$B(E2)_{NPSM}$$^{b}$&	 & $B(E2)_{sdg-IBM}$$^{c}$	\\\hline
   $2_1^+ \rightarrow 0_1^+$	&	  &  0.563 $\pm$ 0.002	&	  &	0.573	&	 & 	0.560 \\
   $4_1^+ \rightarrow 2_1^+$	&	  &  0.819 $\pm$ 0.038	&	  &	0.804	&	 & 	0.810 \\
   $6_1^+ \rightarrow 4_1^+$	&	  &  0.980 $\pm$ 0.09	&	  &	0.844	&	 & 	0.883 \\
   $0_2^+ \rightarrow 2_1^+$	&	  &  0.208 $\pm$ 0.009	&	  &	0.016	&	 & 	0.071 \\
   $2_3^+ \rightarrow 4_1^+$	&	  &  0.095 $\pm$ 0.028	&	  &	0.0	   &	 & 	0.033 \\
   $2_2^+ \rightarrow 2_1^+$	&	  &  0.034 $\pm$ 0.007	&	  &	0.001	&	 & 	0.073 \\
   $2_2^+ \rightarrow 0_1^+$	&	  &  0.015 $\pm$ 0.0009	&	  &	0.0	&	 & 	0.012 
\\  \hline \hline
\end{tabular}\\
\footnotesize{$^{\rm a}$ Reference \cite{PhysRevC.48.461, nndc}.}
\footnotesize{$^{\rm b}$ Present calculation.}
\footnotesize{$^{\rm c}$ Reference \cite{JSimilarity}.}
\label{tab_NdBE2}
\end{table}

From above analysis, one can see that although the number of the states in M-scheme is larger than those in the J-scheme, the $cpu$ time is much smaller in M-scheme than those in J-scheme. And it is challenge  to use the M-scheme to study the nuclei with more valance nucleons.

\section{Acknowledgements}
This work was supported by the Natural Science Foundation of China (11475091, 11875171, 11875158, 11175078, 10935001, 11675071), the Natural Science Foundation for Talent Training in Basic Science (J1103208), the US National Science Foundation (OCI-0904874 and ACI-1516338), US Department of Energy (DE-SC0005248) and the LSU-LNNU joint research program (9961) is acknowledged.
\bibliography{citedpapers}

\end{document}